\documentclass{article}

\usepackage{arxiv}

\usepackage[utf8]{inputenc} % allow utf-8 input
\usepackage[T1]{fontenc}    % use 8-bit T1 fonts
\usepackage{hyperref}       % hyperlinks
\usepackage{url}            % simple URL typesetting
\usepackage{booktabs}       % professional-quality tables
\usepackage{amsfonts}       % blackboard math symbols
\usepackage{nicefrac}       % compact symbols for 1/2, etc.
\usepackage{microtype}      % microtypography
\usepackage{lipsum}
\newcommand{\vect}[1]{\boldsymbol{#1}}
\usepackage{lineno,hyperref}
\usepackage{amsmath,graphicx}
\usepackage{enumitem}
\usepackage{cleveref}
\usepackage{mathrsfs}
\usepackage{thm-restate}
\usepackage{mathtools,amssymb,lipsum, nccmath}
\usepackage{cite}
\usepackage{mathabx}
\usepackage[linesnumbered,ruled,vlined]{algorithm2e}

\newcommand{\tbtm}[3]{
	\left[
	\begin{array}{cc}
		#1  & #3 \\
		#3  & #2
	\end{array}
	\right]
}
\let\ACMmaketitle=\maketitle
\renewcommand{\maketitle}{\begingroup\let\footnote=\thanks \ACMmaketitle\endgroup}

\DeclareMathOperator*{\argmin}{arg\,min}

\usepackage{multirow}
\newtheorem{theorem}{Theorem}
\title{Fast  roughness  minimizing image restoration under mixed Poisson-Gaussian noise\footnote{Submitted to IEEE-TIP}}

\author{
  Manu Ghulyani \\
 Department of Electrical Engineering\\
 Indian Institute of Science\\
 Bengaluru, India \\
 \texttt{manug@iisc.ac.in} \\
   \And
 Muthuvel Arigovindan \\
  Department of Electrical Engineering\\
  Indian Institute of Science\\
  Bengaluru, India \\
  \texttt{mvel@iisc.ac.in} \\
}

\begin{document}
\maketitle

\begin{abstract}
	Image acquisition in many biomedical imaging modalities is
	corrupted by Poisson noise followed by additive Gaussian
	noise.   While total variation and related regularization 
	methods for solving biomedical inverse problems are known
	to yield high quality reconstructions in most situations,  
	such methods mostly
	use log-likelihood of either Gaussian or Poisson noise models, 
	and rarely use mixed Poisson-Gaussian (PG) noise model.  
	The work of Chouzenoux et al. 
	deals with exact PG likelihood and
	total variation regularization. { This method  adapts the primal-dual approach
		involving gradients steps on the PG log-likelihood, with step size
		limited by  the inverse of the Lipschitz constant of the gradient}.
	This leads to limitations in the convergence speed.
	Although     ADMM methods  do not have such step size
	restrictions,  ADDM has never been applied
	for this problem,  for the possible reason that  PG log-likelihood
	is quite complex. 
	In this paper,  we develop an ADMM based optimization for
	roughness minimizing image restoration  under PG log-likelihood.
	We achieve this by first developing a novel iterative method
	for computing the proximal solution of PG log-likelihood,   
	deriving the  termination conditions for this iterative method,
	and then integrating into a provably convergent
	ADMM scheme. The
	effectiveness of the proposed methods is demonstrated using
	restoration examples.
\end{abstract}

\keywords{
	Image Restoration, Maximum likelihood
	estimator (MLE), Alternating direction method of multipliers
	(ADMM), Poisson-Gaussian noise, Total variation, Regularization}

%\linenumbers

\section{Introduction}
The restoration of images from  blur  and noise is an important problem with 
applications in microscopy \cite{jezierska2014approach},\cite{zhu2012faster}, 
\cite{marnissi2016fast} astronomy\cite{benvenuto2008study}, \cite{snyder1995compensation} 
and other sciences. Image restoration is 
often posed as MAP estimation problems  constructed using  a wide variety of 
assumptions on prior probability of the underlying image,  and conditional 
probability of measured pixel values given the degradation model. Various priors such as 
sparsity in  wavelet domain \cite{figueiredo2003algorithm}, sparsity in space 
domain  \cite{zhu2012faster}, sparsity of spatial derivatives \cite{chambolle2004algorithm}, 
\cite{rudin1992nonlinear} promote  different  types of  structures in the recovered image.
The cost functionals corresponding to derivative-based priors are known as
total variation functionals \cite{rudin1992nonlinear}.  Initially, first order image derivative was 
used to construct such functionals, in which case the functionals are known as first order
total variation functionals.   Then it has been demonstrated that second-order total
variation functionals built using second order derivatives yield better reconstruction
quality \cite{lefkimmiatis2012hessian}; in particular, the use of such functionals avoid
the staircase effect \cite{lysaker2006iterative} caused by first-order functionals.
% Currently,  total variation (TV) based regularization functionals,  especially the second order
% TV,  are the most preferred form of regularization functionals.  This is confirmed by the
% fact that there has been significant improvements on this particular category of functionals [refs].

The data fitting (fidelity) term is essentially the negative logarithm of the conditional probability
of the measured pixel value given the ideal measurable pixel value;  it  is also 
known as likelihood function, and it indeed accounts for the probability distribution of the
random process that generates the noise in the measurement device.
Most commonly used  data fitting models are Gaussian and Poisson because of their
simplicity in computation and modelling.  However, 
the degradation caused in image capturing devices such as EM-CCD or CMOS devices is
appropriately modeled by a Poisson process signifying the photon counting followed by 
the additive Gaussian noise accounting for thermal errors \cite{snyder1995compensation}.
This motivates image restoration under the mixed Poisson-Gaussian model (PG) model.
This model is especially relevant in case of biological \cite{bajic2016blind} and 
astronomical imaging \cite{snyder1995compensation}.
This paper aims to develop a faster and practical algorithm for image 
restoration using the MLE-based data fitting  term involving exact Poisson-Gaussian 
(PG) likelihood and the class of convex regularization functionals that have a closed
form proximal solution \cite{bertsekas1999nonlinear}.

Most of the published works in image restoration involving PG likelihood function employ 
some approximation
such as  Generalized Anscombe Transform (GAST) \cite{anscombe1948transformation}, \cite{GAST1,donoho1993,marnissi2016fast} or (shifted) Poisson approximation
\cite{chakrabarti2012image,marnissi2017variational}.  GAST, a variance stabilizing transform, is a non-linear (square root) transform which is applied on the measurements in order that the noise statistics in the measurements are well approximated by a Guassian distribution. The approximation is closer when the mean of the Poisson random variable is high \cite{zhang2007multiscale}.  In \cite{GAST1}, a two stage de-noising approach is developed using GAST. In the first stage, the measured data is applied with Anscombe Transform to `gaussianize' the data, and in the second stage a sparsity driven iterative algorithm is employed to obtain the final solution. 
In shifted Poisson approximation \cite{chakrabarti2012image,marnissi2017variational}, the measurements are added (shifted) with variance of Gaussian random variable, in order that noise in the result is approximated by a Poisson distribution. This approximation performs well when the Gaussian noise variance is low.
The work of Marnissi et al. 
\cite{marnissi2017variational} also considers GAST along with shifted Poisson under
Bayesian framework. 
This approach relies on joint estimation of the signal and the regularization parameter. 
Gao et al. \cite{gao2018bayesian} presented an interesting approach which models the Poisson-Gaussian likelihood as a mixture of Gaussians. The prior considered was Markov Random field prior. The de-noising/de-blurring problem in the approach was formulated as a joint estimation of prior parameters, likelihood parameters and the image variable. 
The restoration results from these algorithms are not as
good as the ones obtained using exact PG likelihood \cite{chouzenoux2015convex},
and there is a scarcity of algorithms considering exact PG likelihood term with TV 
regularization. Furthermore,  the methods  that use the exact PG likelihood
\cite{benvenuto2008study, chouzenoux2015convex}
either has issues in convergence or do not use total variation based regularization
functionals.  Specifically, the  scaled gradient algorithm \cite{benvenuto2008study} 
does not have  any convergence guarantees, and also it does not consider any regularization term. 

Chouzenoux et al. \cite{chouzenoux2015convex} have proposed a rigorous
and general approach for image restoration under PG noise model with total variation.
Its generality stems from the fact that, the approach can be extended to any regularization. However,  
the the  step-size of the   iterative method is restricted to be lower than  the inverse of the 
Lipschitz constant of the gradient of  the log-likelihood functional. Hence,  the convergence is
typically slow.  

ADMM based methods are attractive in the sense that they do not face any  limitation in the step-size,
and hence are typically faster than methods that use gradient-based stepping.
An ADMM method applied on a composite cost 
functional is comprised of a series of minimization steps that cycles through the sub-functionals of the composite cost functional.    The original framework  \cite{eckstein1992douglas} requires that the sub-functionals in
each cycle has to be minimized exactly.  This framework has been used
for image restoration
under  Poisson noise  and under Gaussian noise, and have been shown to be faster than other state-of-the-art methods \cite{figueiredo2010restoration,afonso2010fast}.
As the PG log-likelihood is complex, exact minimization of corresponding sub-functional is not possible in the present 
problem.  Fortunately, a recently proposed modified ADMM framework allows inexact minimization of the sub-functionals
in the ADMM cycles \cite{eckstein2017approximate}.
In this paper,  we adapt this framework for the problem of image restoration using 
convex regularization functionals under the PG noise model.   Our contributions are the following:
\begin{itemize}
	\item
	We propose   an iterative method for minimizing the sub-functional corresponding to PG log-likelihood,
	with proof of convergence.
	\item
	We derive  termination conditions for the above-mentioned iterative scheme such that
	it can be integrated into the modified ADMM framework of Eckstein et al.  \cite{eckstein2017approximate}.
\end{itemize}

In Section \ref{sec:admm},  we will review the ADMM applied on the problem of roughness
minimizing image restoration (subsections  \ref{sec:admmcf}  and \ref{sec:admmintro}).
We will also review  the modified framework of Yao and Eckstein, which enables  solving image restoration under 
Poisson-Gaussian noise model (referred as PG image restoration hereafter) 
by  means of ADMM approach.  Further, we  identify the   computational
problems to be solved for making the modified ADMM framework applicable to PG image restoration
(subsection \ref{sec:admmpgch}).  Section \ref{sec:pgprox} solves these computational problems,
which is the primary contribution of this paper.
Experimental results are given  in Section \ref{sec:majhead}.   This work is an extension of the
work we presented in \cite{islpgtv_isbi2018}, where we proposed the computational algorithm
without convergence proof. 
%In section 2, we describe the Poisson-Gaussian noise model. In section 
%3, we describe the scheme of c ADMM method.  In  section 4, we develop our ADMM based 
%algorithm along with details of all the sub-problems also the iterations involved for EM and
% Scaled-gradient method for  PG noise model . In section 5, we give the results showing the practicability 
%and speed.
%

\section{Roughness minimizing image restoration under Poisson-Gaussian Noise by ADMM}
\label{sec:admm}

\subsection{The cost function}
\label{sec:admmcf}

Let $\vect{g}$ and $\vect{m}^{\prime}$  be the vectors containing the pixels of 
original and measured images  respectively in a scanned form.
Let $\vect{H}$ be matrix equivalent of blurring.   The  measurement
vector $\vect{m}^{\prime}$  differs from the ideal measurement  $\vect{H}\vect{g}$
by noise.   Let ${F}_{\scriptscriptstyle M}(\vect{H}\vect{g}, \vect{m}^{\prime})$  be the 
data-fitting cost functional, which is essentially the negative log of the likelihood
of the noise process.   In other words, 
${F}_{\scriptscriptstyle M}(\vect{H}\vect{g}, \vect{m}^{\prime})=
-\log p_{\scriptscriptstyle M}(\vect{m}^{\prime}|(\vect{Hg}))$, where
\begin{equation}
\label{eq:pyhg}
p_{\scriptscriptstyle M}(\vect{m}^{\prime}|(\vect{Hg}))=\prod_n
p_{\scriptscriptstyle M}((\vect{m}^{\prime})_{n}|(\vect{Hg})_{n}),
\end{equation}
with $p_{\scriptscriptstyle M}((\vect{m}^{\prime})_{n}|(\vect{Hg})_{n})$ denoting the
likelihood for  $(\vect{Hg})_{n}$ being the ideal $n$th  pixel given  
$(\vect{m}^{\prime})_{n}$ as the $n$th  measured pixel.  When the noise is
assumed to be Gaussian,   it is given by
$$p_{\scriptscriptstyle M}((\vect{m}^{\prime})_{n}|(\vect{Hg})_{n})=\frac{1}{\sqrt{2\pi\sigma^2}}
\exp\left(-\frac{1}{\sigma^2}\left((\vect{m}^{\prime})_{n}-(\vect{Hg})_{n}\right)^2\right),$$
where $\sigma^2$ is the noise variance. 
When the noise is assumed to be Poisson,  it becomes
$$p_{\scriptscriptstyle M}((\vect{m}^{\prime})_{n}|(\vect{Hg})_{n})=
\exp(-(\vect{Hg})_{n})\frac{(\vect{Hg})_{n})^{(\vect{m}^{\prime})_{n}}}
{((\vect{m}^{\prime})_{n})!}.$$
The most realistic form of noise model,  which is the focus of this paper, is the mixed
Poisson-Gaussian noise model.  In this case, $\vect{m}^{\prime}$  and $\vect{Hg}$
are related as given below,
\begin{equation} \label{eq:pgform}
%\begin{split}
(\vect{\vect{m}^{\prime}})_{n}  = \alpha \mathcal P((\vect{Hg})_{n})+ \mathcal N(c,\sigma^2),\\
%& = \frac{1}{2} \pi r^2
%\end{split}
\end{equation}
where  $\mathcal P(\cdot)$ denotes the Poisson process, and $\mathcal N(c,\sigma^2)$ is Gaussian process
with mean $c$ and variance $\sigma^2$. Note that we consider $c=0$ in this paper. The corresponding 
likelihood is given by
\begin{align} \label{eq:pgl1}
& p_{\scriptscriptstyle M}((\vect{m}^{\prime})_{n}|(\vect{Hg})_{n})   =  \\ \nonumber&
\sum_{p=0}^{\infty}\frac{e^{-(\vect{Hg})_{n}}}{\sqrt{2\pi \sigma ^2}}
\frac{((\vect{Hg})_{n})^p}
{p!} \exp\left(-\frac{((\vect{m}^{\prime})_{n}-\alpha p-c)^2}{2\sigma^2}\right).
\end{align}  

Next, let $\vect{D} = \left[\vect{D}_{xx}^T\;\vect{D}_{yy}^T\; \vect{D}_{xy}^T\right]^T$,
where $\vect{D}_{xx}$, $\vect{D}_{yy}$ and $\vect{D}_{xy}$ denote matrix 
equivalent of convolving the image with filters corresponding to derivative operators
$\frac{\partial^2}{\partial x^2}$, $\frac{\partial^2}{\partial y^2}$, and 
$\frac{\partial^2}{\partial x\partial y}$ respectively.  This means $\vect{D}_{xx}, \vect{D}_{xy}$ and $\vect{D}_{yy}$ are block circulant matrices with circulant blocks corresponding to 2-D convolution with periodic boundary conditions for the filters $[-1,2,-1]$,$\left[
\begin{array}{cc}
1  & -1 \\
-1  & 1
\end{array}
\right]$ and $\left[\begin{array}{c}
-1    \\
2  \\
-1 
\end{array}
\right]$ respectively. Let $\mathbf{P}_i$ denote the matrix with 
$1$s at positions $(i,1), (i+N,2), (i+2N,3)$
and zeros everywhere else.  Further,  let $\mathcal{E}(\mathbf{v})$ be the operator defined
for the vector $\mathbf{v}=[v_1,v_2,v_3]^T\in\mathbb{R}^3$  that returns vector of Eigen values of the 
matrix 
$\tbtm{v_1}{v_2}{v_3}$.   Then most second order derivative based roughness  functionals
fall under category of Hessian-Schatten norm \cite{6403545}, which can be expressed as given below,
\begin{equation}
\label{eq:fddef1}
F_{\scriptscriptstyle D}(\vect{D}\vect{g}) = \sum_{i=1}^N
\left\|\mathcal{E}(\vect{P}_i^T\vect{D}\vect{g})\right\|_q,
\end{equation}
where $q$ is a parameter in the range $[1,\infty]$.  This functional is computationally least
expensive when $q=1,2$. When $q=2$, this form becomes the well-known total variation
functional.  When $q=1$, the norm is known as the nuclear norm,  which has been reported
to yield better results. 

With these definitions,  the roughness minimizing image restoration amounts to computing
the minimum of the following  cost,
\begin{equation}
\label{eq:fullcost1p}
F(\vect{g}) = {F}_{\scriptscriptstyle M}(\vect{H}\vect{g}, \vect{m}^{\prime}) + 
\lambda  F_{\scriptscriptstyle D}(\vect{D}\vect{g}) + F_B(\vect{g})
\end{equation}
where $\lambda$ is the regularization parameter, and $F_B(\vect{g})$   is the indicator 
function for imposing
bound constraint on the image pixel values. With $u^{\prime}$ (a positive real number)
denoting largest pixel value that can be allowed in the restoration,  $F_B(\vect{g})$
can be written as
\begin{equation}
\label{eq:fbdef}
F_B(\vect{g}) = \sum_{i=1}^N 
\begin{cases}
0 & \mbox{if}   \; 0 \le (\vect{g})_i \le u^{\prime}, \\
\infty, & \mbox{otherwise.}
\end{cases}
\end{equation}
Next, we propose to modify 
${F}_{\scriptscriptstyle M}(\vect{Hg}, \vect{m}^{\prime})$
as given below:
\begin{equation}
\label{eq:fmdef2}
\bar{F}_{\scriptscriptstyle M}(\vect{H}\vect{g}, \vect{m}^{\prime}) = 
\sum_{i=1}^N
\begin{cases}
-\ln(p_{\scriptscriptstyle M}((\vect{m}^{\prime})_i|(\vect{Hg})_i)) &\\\;\;\;\;\;\;\;\; \mbox{if} \;
(\vect{Hg})_i \in [l,u], \\
\infty, & \mbox{otherwise}.
\end{cases}
\end{equation}
This is equivalent to imposing the constraint that the components of  $\vect{Hg}$ stay
within the bound $[l,u]$. First note that $0\le l\le u$. Secondly, $l$ can be set to zero and $u$ to $|| \vect{H}||_1 u' $. The second result is obtained by a straight forward application of Holder's inequality. Although,  the above bounds clearly redundant because of the bound constraint
on $\vect{g}$,  this helps to make to ADMM iteration well-behaved.    This will be explained
later in Section IV.B.  With this modification, the overall cost to be minimized is given by
\begin{equation}
\label{eq:fullcost1}
F(\vect{g}) = \bar{F}_{\scriptscriptstyle M}(\vect{H}\vect{g}, \vect{m}^{\prime}) + 
\lambda  F_{\scriptscriptstyle D}(\vect{D}\vect{g}) + F_B(\vect{g}).
\end{equation}

\subsection{The ADMM algorithm}
\label{sec:admmintro}

The first step in developing the ADMM algorithm is to consider the following minimization
problem:
\begin{eqnarray}
\label{eq:fullcostminc}
(\vect{g}_{opt},\vect{m}_{opt},\vect{d}_{opt},\vect{b}_{opt}) = \\ \nonumber
\argmin_{(\vect{g},\vect{m},\vect{d},\vect{b})} 
\bar{F}_{\scriptscriptstyle M}(\vect{m}, \vect{m}^{\prime}) + 
\lambda  F_{\scriptscriptstyle D}(\vect{d}) +Â F_B(\vect{b}) 
\\ \nonumber s.t.  \;\; \vect{H}\vect{g} = \vect{m},  \vect{D}\vect{g} = \vect{d},  \vect{g} = \vect{b}.
\end{eqnarray}
Clearly,  $\vect{g}_{opt}$ obtained from solving the above optimization problem is also
the minimum of the function $F(\vect{g})$ given in the equation  \eqref{eq:fullcost1}.

The ADMM method is similar to augmented Lagrangian approach
developed for constrained optimization problems \cite{bertsekas1999nonlinear}.  
The first step is to write the augmented
Lagrangian function of the above constrained optimization problem.  To this end, 
we define the following:
\begin{align}
\label{eq:auglagm}
L_{\scriptscriptstyle M}(\vect{g},\vect{m},\hat{\vect{m}}, \vect{m}^{\prime})  & = 
\medmath{\bar{F}_{\scriptscriptstyle M}(\vect{m}, \vect{m}^{\prime}) + 
	\frac{\beta}{2} \left\|\vect{H}\vect{g}-\vect{m}\right\|^2_2 - \hat{\vect{m}}^T(\vect{H}\vect{g}-\vect{m})} \\ 
\label{eq:auglagd}
L_{\scriptscriptstyle D}(\vect{g},\vect{d},\hat{\vect{d}})  & =  \lambda F_{\scriptscriptstyle D}(\vect{d}) + 
\frac{\beta}{2} \left\|\vect{D}\vect{g}-\vect{d}\right\|^2_2 - \hat{\vect{d}}^T(\vect{D}\vect{g}-\vect{d}) \\
\label{eq:auglagb}
L_{\scriptscriptstyle B}(\vect{g},\vect{b},\hat{\vect{b}})  & =  F_{\scriptscriptstyle B}(\vect{b}) + 
\frac{\beta}{2} \left\|\vect{g}-\vect{b}\right\|^2_2 - \hat{\vect{b}}^T(\vect{g}-\vect{b}) \end{align}
With these definitions, the augmented Lagrangian of the problem of equation
\eqref{eq:fullcostminc} is define as
\begin{eqnarray}
\label{eq:auglag}
&L(\vect{g},\vect{m},\hat{\vect{m}},\vect{d},\hat{\vect{d}},\vect{b},\hat{\vect{b}},\vect{m}^{\prime}) 
=\\ \nonumber &
L_{\scriptscriptstyle M}(\vect{g},\vect{m},\hat{\vect{m}},\vect{m}^{\prime}) +  
L_{\scriptscriptstyle D}(\vect{g},\vect{d},\hat{\vect{d}}) +
L_{\scriptscriptstyle B}(\vect{g},\vect{b},\hat{\vect{b}}).
\end{eqnarray}
Here, the variables $(\hat{\vect{m}},\hat{\vect{d}},\hat{\vect{b}})$ are called
Lagrange's multipliers.  With this definition,   the ADMM method involves series of
minimizations on  $L(\vect{g},\vect{m},\hat{\vect{m}},\vect{d},\hat{\vect{d}},
\vect{b},\hat{\vect{b}}, \vect{m}^{\prime})$, where each minimization is done with respect to one of the
variables in the set  $(\vect{g},\vect{m},\vect{d},\vect{b})$.  Selection of the variable
for minimization,   cycles through the list  $(\vect{g},\vect{m},\vect{d},\vect{b})$,
and each cycle is considered as one step  of the ADMM iteration.  
In other words, if $k$ is the iteration index,  the update from 
$(\vect{g}^{(k)},\vect{m}^{(k)},\hat{\vect{m}}^{(k)},\vect{d}^{(k)}$,
$\hat{\vect{d}}^{(k)},\vect{b}^{(k)},\hat{\vect{b}}^{(k)})$
to 
$(\vect{g}^{(k+1)},\vect{m}^{(k+1)},\hat{\vect{m}}^{(k+1)},\vect{d}^{(k+1)},
\hat{\vect{d}}^{(k+1)},\vect{b}^{(k+1)},\hat{\vect{b}}^{(k+1)})$
can be expressed in term of the following steps: 
\begin{flalign}
\label{eq:admmg}
&Step\;1:  \;\;\vect{g}^{(k+1)}   =  \\ \nonumber & \argmin_{\vect{g}}
L(\vect{g},\vect{m}^{(k)},\hat{\vect{m}}^{(k)},
\vect{d}^{(k)},\hat{\vect{d}}^{(k)},\vect{b}^{(k)},\hat{\vect{b}}^{(k)},\vect{m}^{\prime}) \\
\label{eq:admmm}
&Step\;2: \;\;\vect{m}^{(k+1)}  = \\ \nonumber&  \argmin_{\vect{m}}
L(\vect{g}^{(k+1)},\vect{m},\hat{\vect{m}}^{(k)},
\vect{d}^{(k)},\hat{\vect{d}}^{(k)},\vect{b}^{(k)},\hat{\vect{b}}^{(k)},\vect{m}^{\prime}) \\
\label{eq:admmd}
&Step\;3: \;\;\vect{d}^{(k+1)}  = \\ \nonumber&   \argmin_{\vect{d}}
L(\vect{g}^{(k+1)},\vect{m}^{(k+1)},\hat{\vect{m}}^{(k)},
\vect{d},\hat{\vect{d}}^{(k)},\vect{b}^{(k)},\hat{\vect{b}}^{(k)},\vect{m}^{\prime}) \\
\label{eq:admmb}
&Step\;4: \;\;\vect{b}^{(k+1)}  = \\ \nonumber&  \argmin_{\vect{b}}
L(\vect{g}^{(k+1)},\vect{m}^{(k+1)},\hat{\vect{m}}^{(k)},
\vect{d}^{(k+1)},\hat{\vect{d}}^{(k)},\vect{b},\hat{\vect{b}}^{(k)},\vect{m}^{\prime}) \\ \nonumber
\label{eq:admmmhu}
\end{flalign}
\begin{flalign}
Step\;5:  \hat{\vect{m}}^{(k+1)} & =   \hat{\vect{m}}^{(k)}-\beta(\vect{H}\vect{g}^{(k+1)}-\vect{m}^{(k+1)}), \\
\label{eq:admmdhu}
\hat{\vect{d}}^{(k+1)} & =   \hat{\vect{d}}^{(k)}-\beta(\vect{D}\vect{g}^{(k+1)}-\vect{d}^{(k+1)}), \\
\label{eq:admmbhu}
\hat{\vect{b}}^{(k+1)} & =   \hat{\vect{b}}^{(k)}-\beta(\vect{g}^{(k+1)}-\vect{b}^{(k+1)})
\end{flalign}
Taking into account the dependency of sub-functionals of 
$L(\vect{g},\vect{m},\hat{\vect{m}},\vect{d},\hat{\vect{d}},
\vect{b},\hat{\vect{b}}, \vect{m}^{\prime})$ on the variables involved in the minimizations,  
Steps $1-4$ can be also
expressed as follows:
\begin{eqnarray}
\label{eq:admmg2}
Step\;1:  \;\vect{g}^{(k+1)}  & = &  \argmin_{\vect{g}}
Q(\vect{g},\vect{m}^{(k)},\vect{d}^{(k)},\vect{b}^{(k)}) \\
\label{eq:admmm2}
Step\;2: \;\vect{m}^{(k+1)} & = & \medmath{ \argmin_{\vect{m}}
	L_{\scriptscriptstyle M}(\vect{g}^{(k+1)},\vect{m},\hat{\vect{m}}^{(k)}, \vect{m}^{\prime})} \\
\label{eq:admmd2}
Step\;3: \;\;\vect{d}^{(k+1)} & = &  \argmin_{\vect{d}}
L_{\scriptscriptstyle D}(\vect{g}^{(k+1)},
\vect{d},\hat{\vect{d}}^{(k)}) \\
\label{eq:admmb2}
Step\;4: \;\;\vect{b}^{(k+1)} & = &  \argmin_{\vect{b}}
L_{\scriptscriptstyle B}(\vect{g}^{(k+1)}, 
\vect{b},\hat{\vect{b}}^{(k)}) 
\end{eqnarray}
where  
\begin{equation}
\label{eq:qdef}
\begin{split}
&Q(\vect{g},\vect{m}^{(k)},\vect{d}^{(k)},\vect{b}^{(k)})  = \\&
\frac{\beta}{2} \left\|\vect{H}\vect{g}-\vect{m}^{(k)}\right\|^2_2 +
\frac{\beta}{2} \left\|\vect{D}\vect{g}-\vect{d}^{(k)}\right\|^2_2  + 
\frac{\beta}{2} \left\|\vect{g}-\vect{b}^{(k)}\right\|^2_2  \\
& - (\hat{\vect{m}}^{(k)})^T(\vect{H}\vect{g}-\vect{m}^{(k)}) 
-  (\hat{\vect{d}}^{(k)})^T(\vect{D}\vect{g}-\vect{d}^{(k)}) \\
&  - (\hat{\vect{b}}^{(k)})^T(\vect{g}-\vect{b}^{(k)}).
\end{split}
\end{equation}
The initialization for the above iteration can be set to zero for entire set
$(\vect{g},\vect{m},\hat{\vect{m}},\vect{d}$, $\hat{\vect{d}},
\vect{b},\hat{\vect{b}})$,  and iteration can be typically terminated based
on the relative change on the required image, i.e., 
$\frac{\left\| \vect{g}^{(k+1)}-\vect{g}^{(k)} \right\|_2}
{\left\|\vect{g}^{(k+1)}\right\|_2}$.
It has been shown by Eckstein et al.  \cite{eckstein1992douglas}   that the above
iteration represented by Steps 1-5 converges to the solution of the problem
\eqref{eq:fullcostminc}---which is the same as the minimum of the original cost $F(\vect{g})$
given in the equation \eqref{eq:fullcost1}---if the following
conditions are satisfied: (1) the sub-functions are closed, which is true in our
case, i.e.,  the functions $\bar{F}_{\scriptscriptstyle M}$, 
$F_{\scriptscriptstyle D}$,  and $F_{\scriptscriptstyle B}$ are closed; 
(2) the minimization denoted in the Steps 1-4 are exact;
(3)  the  matrix  obtained by vertically augmenting the matrices involved in the
equality constraints (equation \eqref{eq:fullcostminc})   should have
full column rank,  which also true in our case since one of the matrices is identity. Note that, a convex function $f$ is called a closed function if every sub-level set ($\{x\in dom(f)|f(x)\le t\}$)  is closed.

The minimization problems represented by  Steps 1, 3,  and 4  are actually single step
minimizations  meaning that,  the solutions can be obtained through specific formulas. 
These 
formulas  are well-known,  and
for the readers' convenience, they are given in Appendix A. 
The  minimization of Step 2 (\cref{eq:admmm2}) can also be solved in single step if 
$p_{\scriptscriptstyle M}(\vect{m}^{\prime}|(\vect{m}))$ either purely Gaussian or
Poisson \cite{afonso2011augmented, figueiredo2010restoration}.
When $p_{\scriptscriptstyle M}(\vect{m}^{\prime}|(\vect{m}))$ corresponds to 
mixed Poisson-Gaussian model,   Step 2 has to be solved iteratively because 
$L_{\scriptscriptstyle M}(\vect{g}^{(k+1)},\vect{m},\hat{\vect{m}}^{(k)}, \vect{m}^{\prime})$
becomes complex to minimize.  This also means that this step cannot be solved exactly,
and hence classic ADMM theory of convergence \cite{eckstein1992douglas}  will not be applicable. 
Hence the modified framework \cite{eckstein2017approximate} has to be used; however,  this modified framework is not
directly applicable,  and it requires solving some computational problems as elaborated in the
following sub-section.   
These problems are   addressed in \cref{sec:pgprox}, which is the main focus of this paper.

\subsection{The issues in implementing ADMM for Poisson-Gaussian noise model}
\label{sec:admmpgch}

As mentioned before, application of classic convergence theory of ADMM requires that 
Steps-1--4 of equation 
\eqref{eq:admmg2}, \eqref{eq:admmm2}, \eqref{eq:admmd2}, and \eqref{eq:admmb2}
has to solved exactly.  We also indicated that the step  2 cannot be solved exactly.
To proceed further,  let 
${L}_{\scriptscriptstyle M,k}(\vect{m}, \vect{m}^{\prime})$ denote the cost to be 
minimized in the  Step 2,  which can be written as
\begin{equation}
\label{eq:lmkdef} 
\begin{split}
L_{\scriptscriptstyle M, k}(\vect{m},  \vect{m}^{\prime})  &
= 
\bar{F}_{\scriptscriptstyle M}(\vect{m}, \vect{m}^{\prime}) + 
\frac{\beta}{2} \left\|\vect{H}\vect{g}^{(k+1)}-\vect{m}\right\|^2_2  \\
& +  (\hat{\vect{m}}^{(k+1)})^T(\vect{H}\vect{g}-\vect{m}).
\end{split}
\end{equation}
This can also be written as
\begin{equation}
\label{eq:lmkdef2}
\begin{split}
&{L}_{\scriptscriptstyle M,k}( \vect{m}, \vect{m}^{\prime} )   =    
\bar{F}_{\scriptscriptstyle M}(\vect{m}, \vect{m}^{\prime}) + 
\frac{\beta}{2} \left\|\vect{m}-\bar{\vect{m}}^{(k+1)}\right\|^2_2 \\
&
\bar{\vect{m}}^{(k+1)} = \vect{H}{\bf g}^{(k+1)}-\frac{1}{\beta}\hat{\vect{m}}^{(k)}
\end{split}
\end{equation}
Note that, in the above equation, 
${F}_{\scriptscriptstyle M}(\vect{m}, \vect{m}^{\prime})=
-\log {p}_{\scriptscriptstyle M}(\vect{m}, \vect{m}^{\prime})$,
where ${p}_{\scriptscriptstyle M}(\vect{m}, \vect{m}^{\prime})$ is given by the equations
\eqref{eq:pgl1}  and  \eqref{eq:pyhg}.
If exact  Poisson-Gaussian model is used for
${p}_{\scriptscriptstyle M}(\vect{m}, \vect{m}^{\prime})$,
there will be no single step minimization solution for this,  and has to be
minimized iteratively.  This will also mean that
${L}_{\scriptscriptstyle M,k}( \vect{m}, \vect{m}^{\prime} )$  cannot be
solved exactly.  
To ensure convergence in this case, 
the modified  ADMM  framework of Eckstein and Yao
\cite{eckstein2017approximate} has to be used.    To write  the required adoption of
this framework for our problem,  we first  re-express the problem given in Step 2 (equation 
\eqref{eq:admmm2})  as given below:
\begin{align}
\label{eq:admmm3a}
Step\;2a: & \;\;  c_{k} =  \left\|\vect{D}\vect{g}^{(k+1)}-\vect{d}^{(k)}\right\|^2_2  + 
\left\|\vect{g}^{(k+1)}-\vect{b}^{(k)}\right\|^2_2\\
\label{eq:admmm3b}
Step\;2b: & \;\;[\vect{m}^{(k+1)}, \pmb{\eta}^{(k+1)}]  =\\ \nonumber&
\mathlarger{\mathlarger{\cal I}}_{[\vect{m}]}\left[L_{\scriptscriptstyle M,k}(
\vect{m},\vect{m}^{\prime}), \bar{\vect{m}}^{(k)}, \vect{w}^{(k)}, c_k\right] \\
\label{eq:admmm3c}
Step\;2c: & \;\; \vect{w}^{(k+1)} =  \vect{w}^{(k)} - \beta\pmb{\eta}^{(k)}  
\end{align}
In step 2a,  we compute the current quadratic constraint error with respect to sub-problems
corresponding to roughness and out-of-bound penalty, which is denoted by $c_k$. 
Step 2b, calls the iterative refinement,  denoted by
$\mathlarger{\mathlarger{\cal I}}_{[\vect{m}]}[L_{\scriptscriptstyle M,k}(\cdot), 
\cdot,\cdot, \cdot]$,   which computes the
successive refinements towards the minimum of 
$L_{\scriptscriptstyle M,k}(\cdot)$  given in the equation \eqref{eq:lmkdef2} 
with respect to the variable $\vect{m}$;
it returns an approximate minimum denoted by $\vect{m}^{(k+1)}$  at 
the attainment of certain termination  conditions.  We will postpone the 
specification of its actual implementation,  and we are now concerned only
on  the conditions
it should satisfy so that the overall algorithm converges.  
We will also assume that it returns the  gradient at $\vect{m}^{(k+1)}$ denoted
by   $\pmb{\eta}^{(k+1)}= \pmb{\eta}(\vect{m}^{(k+1)},\vect{m}^{\prime})$.
The argument, $\bar{\vect{m}}^{(k)}$,  passed to the iterator 
$\mathlarger{\mathlarger{\cal I}}_{[\vect{m}]}[L_{\scriptscriptstyle M,k}(\cdot), 
\cdot, \cdot, \cdot]$ signifies the fact that $L_{\scriptscriptstyle M,k}(\cdot)$
depends on the current iteration index in terms of $\bar{\vect{m}}^{(k)}$
(equation  \eqref{eq:lmkdef2}).  
The other inputs that are not part of the function
$L_{\scriptscriptstyle M,k}(\cdot)$,  namely   $\vect{w}^{(k)}$  and $c_k$
are used to test the termination condition for 
$\mathlarger{\mathlarger{\cal I}}_{[\vect{m}]}[]$. The vector  $\vect{w}^{(k)}$
is essentially an accumulation of past gradients of 
$L_{\scriptscriptstyle M,k}(\cdot)$
with respect to $\vect{m}$.  For $k=0$, this vector can be initialized  to zero.
Here, it is clear that the termination condition for the inner iteration 
$\mathlarger{\mathlarger{\cal I}}_{[\vect{m}]}[L_{\scriptscriptstyle M,k}(\cdot), 
\cdot,\cdot, \cdot]$ is also dependent  on the state of the outer iteration
(ADMM loop) because  $\vect{w}^{(k)}$  and $c_k$ are $k$-dependent.

Note that now the overall algorithm is nesting of two iterations where the 
outer one is the classic ADMM loop,
and inner one is $\mathlarger{\mathlarger{\cal I}}_{[\vect{m}]}
[L_{\scriptscriptstyle M,k}(\cdot), 
\cdot, \cdot]$.  For each value of $k$, which is the iteration index for
the outer loop,   
$\mathlarger{\mathlarger{\cal I}}_{[\vect{m}]}
[L_{\scriptscriptstyle M, k}(\cdot), \cdot,
\cdot, \cdot]$   works on $k$-dependent minimization problem because of the fact
that $L_{\scriptscriptstyle M,k}()$, as a function of $\vect{m}$,
is dependent on $\bar{\vect{m}}^{(k+1)} = \vect{H}{\bf g}^{(k+1)}-\frac{1}{\beta}\hat{\vect{m}}^{(k)}$.
Let $\{\vect{m}^{(k)}_l\}_{l=0,1,2,\ldots}$ be the sequence of iterates generated
by  $\mathlarger{\mathlarger{\cal I}}_{[\vect{m}]}[L_{\scriptscriptstyle M, k}(\cdot), 
\cdot, \cdot,\cdot]$   towards the minimum of $L_{\scriptscriptstyle M,k}()$ with respect to $\vect{m}$.
At the attainment of termination condition,  the algorithm makes the assignment
$\vect{m}^{(k+1)} = \vect{m}^{(k)}_l$.    
Eckstein and Yao \cite{eckstein2017approximate}
have given two termination conditions
to be used inside   
$\mathlarger{\mathlarger{\cal I}}_{[\vect{m}]}
[L_{\scriptscriptstyle M, k}(\cdot), 
\cdot, \cdot,\cdot]$    
such that  the overall ADMM iteration converges to the minimum 
of the problem  given in the equation \eqref{eq:fullcostminc}. These are given below:
\begin{itemize}
	\item
	{\bf Condition 1}: $\|\pmb{\eta}^{(k)}_l\|_2< \theta_k$ where  $\{\theta_k\}$ is a sequence
	of positive real numbers that is summable,  i.e., $\sum_{k=0}^{\infty}\theta_k<\infty$, and 
	$\pmb{\eta}^{(k)}_l$ is the sub-gradient of $L_{\scriptscriptstyle M,k}(\cdot)$
	at $\vect{m}= \vect{m}^{(k)}_l$. \label{condition1}
	\item
	{\bf Condition 2}: $\frac{2|\langle\vect{w}^{(k)}-\vect{m}^{(k)}_l,
		\vect{\pmb{\eta}^{(k)}_l}\rangle|+||\pmb{\eta}^{(k)}_l||_2^2}
	{c_k+ ||\vect{Hg}^{(k+1)}-\vect{m}^{(k)}_l||_2^2}<\rho<1$ for some real number $\rho$.
\end{itemize}
Clearly,   both conditions imply that $\|\pmb{\eta}^{(k)}_l\|_2$  should decrease that as $k$ increases
and hence the number of iterations  in
$\mathlarger{\mathlarger{\cal I}}_{[\vect{m}]}[L_{\scriptscriptstyle M,k }(\cdot), 
\cdot, \cdot,\cdot]$ should increase as $k$ increases.

To construct   converging algorithm for image restoration under exact Poisson-Gaussian model
using this framework,  we need to address 
two problems, which will be the focus of the next section:
\begin{itemize}
	\item
	Find alternative conditions for {\bf Condition 1} and {\bf Condition 2} to accommodate the fact
	that the gradient $\pmb{\eta}(\vect{m},\vect{m}')$ can never be computed exactly since 
	$L_{\scriptscriptstyle M,k}(\cdot)$ will have infinite summations.
	\item
	Construct a converging algorithm for 
	$\mathlarger{\mathlarger{\cal I}}_{[\vect{m}]}[L_{\scriptscriptstyle M,k}(\cdot), \cdot,\cdot, \cdot]$
	such that these conditions can be met.
\end{itemize}

\section{Solving the data-fitting sub-problem}
\label{sec:pgprox}

The minimum of the data-fitting     Lagrangian, 
${L}_{\scriptscriptstyle M,k}( \vect{m}, \vect{m}^{\prime} )$   given in the equation
\eqref{eq:lmkdef2}
is called the proximal of $\bar{\vect{m}}^{(k+1)}$ to 
$\bar{F}_{\scriptscriptstyle M}(\vect{m}, \vect{m}^{\prime})$.  The goal here is to derive
a converging iterative algorithm to compute this minimum.  In the first subsection,
we derive necessary results for constructing the iterative algorithms.
In the next subsection, we construct  two iterative methods for minimizing 
${L}_{\scriptscriptstyle M,k}( \vect{m}, \vect{m}^{\prime})$  by using well-known
general schemes namely damped-Newton method, and majorization-minimization
method.  We also prove convergence for the damped-Newton method.  
In the last subsection, we derive modified termination conditions that need to be
imposed on these iterations, so that, the overall ADMM iteration converges.

\subsection{Analysis of data-fitting Lagrangian}

\subsubsection{\underline{The basic log-likelihood ${F}_{\scriptscriptstyle M}(\vect{m}, \vect{m}^{\prime} )$}} 

To minimize ${L}_{\scriptscriptstyle M,k}( \vect{m}, \vect{m}^{\prime} )$,  
we need its derivatives.   The main complexity in
the above function is in   $\bar{F}_{\scriptscriptstyle M}(\vect{m}, \vect{m}^{\prime} )$
defined in the equation  \eqref{eq:fmdef2}, 
which is an extension of
${F}_{\scriptscriptstyle M}(\vect{m}, \vect{m}^{\prime} )
=-\log \prod_n
p_{\scriptscriptstyle M}((\vect{m}^{\prime})_{n}|(\vect{m})_{n})$
where  $p_{\scriptscriptstyle M}((\vect{m}^{\prime})_{n}|(\vect{m})_{n})$
can be re-written from the equation \eqref{eq:pgl1}  as given below:
\begin{align} \label{eq:pglm1}
& p_{\scriptscriptstyle M}((\vect{m}^{\prime})_{n}|(\vect{m})_{n})   =  \\ \nonumber&
\sum_{p=0}^{\infty}\frac{e^{-(\vect{m})_{n}}}{\sqrt{2\pi \sigma ^2}}
\frac{((\vect{m})_{n})^p}
{p!} \exp\left(-\frac{((\vect{m}^{\prime})_{n}-p)^2}{2\sigma^2}\right).
\end{align}  
We will need the derivatives of ${F}_{\scriptscriptstyle M}(\vect{m}, \vect{m}^{\prime} )$ for
constructing the derivative expressions for ${L}_{\scriptscriptstyle M,k}( \vect{m}, \vect{m}^{\prime} )$.
The first derivative of ${F}_{\scriptscriptstyle M}(\vect{m}, \vect{m}^{\prime})$ 
has been given in
\cite{chouzenoux2015convex}, which is expressed below:
\begin{equation}
\label{eq:fmfd}
\medmath{\pmb{\gamma}_1(\vect{m}, \vect{m}^{\prime}) }=
\medmath{\nabla_{\vect{m}}({F}_{\scriptscriptstyle M}(\vect{m}, \vect{m}^{\prime}))} =\medmath{
	\vect{1}- (\vect{s}(\vect{m}, \vect{m}^{\prime}-1))\oslash(\vect{s}(\vect{m}, \vect{m}^{\prime})),}
\end{equation}
where $\vect{1}$ denotes the vector of $1$'s,
$\oslash$ denotes the element-wise division of the vectors,   and
\begin{equation}
\label{eq:sexp}
\vect{s}(\vect{a},\vect{b})=\sum_{j=0}^{\infty}\frac{\vect{a}^{.j}}
{j!}\exp\left[{\frac{-(\vect{b}-\alpha j)^2}{2\sigma^2}}\right].
\end{equation}
In the above expression, $()^{.j}$ denotes the element-wise
powering of its vector argument. Next, 
note that,  since ${F}_{\scriptscriptstyle M}(\vect{m}, \vect{m}^{\prime})$
has no dependence among the component of $\vect{m}$,  its Hessian is
a diagonal matrix.  Let  $\bar{\nabla}_{\vect{m}}^2$ denote the operator giving
the diagonal elements of the Hessian.  Result of this operation
on  ${F}_{\scriptscriptstyle M}(\vect{m}, \vect{m}^{\prime})$ can be expressed
as  \cite{chouzenoux2015convex}
\begin{equation}
\label{eq:fmsd}
\begin{split}
&\pmb{\gamma}_2(\vect{m}, \vect{m}^{\prime})  = 
\bar{\nabla}_{\vect{m}}^2({F}_{\scriptscriptstyle M}(\vect{m}, \vect{m}^{\prime}))  \\
&   = 
\medmath{
	[\vect{s}^{.2}(\vect{m}, \vect{m}^{\prime}-\vect{1})-
	\vect{s}(\vect{m}, \vect{m}^{\prime})
	\vect{s}(\vect{m}, \vect{m}^{\prime}-\vect{2})]\oslash
	[\vect{s}^{.2}(\vect{m},\vect{m}^{\prime})],}
\end{split}
\end{equation}
where  $\vect{2} \in \mathbb{R}^N$ denotes vector of $2$'s.

From the expressions given above, it is clear that both first and second derivatives
need  approximation,  since they involve infinite summations. 
For an approximation of this expression, we use the approach as followed by 
Chouzenoux et al., \cite{chouzenoux2015convex}. 
$\vect{s}(\vect{a},\vect{b})$  is approximated
by $\vect{s}_\Delta(\vect{a},\vect{b})$ which is defined as:
\begin{equation}
\label{eq:sexp2}
\vect{s}_\Delta(\vect{a},\vect{b})
=\exp(\frac{\vect{b}^{\cdot 2}}{-2\sigma^2})+\sum\limits_{max(1,[\vect{n}^*-\frac{\Delta \sigma}{\alpha}])}^{[\vect{n}^*+\frac{\Delta\sigma}{\alpha}]}
\frac{\vect{a}^{.n} \exp{\frac{(\vect{b}-\alpha n)^2}{-2\sigma^2} }}{n!} \;\; 
\end{equation}
Here, $n_i^*$ is the term which maximizes $\frac{{a_i}^{.n} \exp{\frac{({b_i}-\alpha n)^2}{-2\sigma^2} }}{n!}$ with respect to  $n$. For each vector component of $a_i$ and $b_i$ we get a value of $n_i^*$. Therefore, $\mathbf{n^*}$ is a vector obtained by stacking all $n_i^*$s. In the above expressions, larger the value of $\Delta$,  lower will be the approximation error.
Here, $[x] $ denotes the greatest integer less than $x$. Using the above approximation, we define the following approximated first and
second derivatives:
\begin{eqnarray}
\label{eq:fmfd2}
%\begin{split}
&\pmb{\gamma}_{1,\Delta}(\vect{m}, \vect{m}^{\prime})  =
\vect{1}- (\vect{s}_\Delta(\vect{m}, \vect{m}^{\prime}-1))\oslash(\vect{s}_\Delta(\vect{m}, \vect{m}^{\prime})),  \\
%\end{split}
\label{eq:fmsd2}
&\pmb{\gamma}_{2,\Delta}(\vect{m}, \vect{m}^{\prime})  =\\ \nonumber&\medmath{
	[\vect{s}^{.2}_\Delta(\vect{m}, \vect{m}^{\prime}-\vect{1})-
	\vect{s}_\Delta(\vect{m}, \vect{m}^{\prime})
	\vect{s}_\Delta(\vect{m}, \vect{m}^{\prime}-\vect{2})]\oslash
	[\vect{s}^{.2}_\Delta(\vect{m},\vect{m}^{\prime})],}
\end{eqnarray}

\subsubsection{\underline{The extended log-likelihood 
		$\bar{F}_{\scriptscriptstyle M}(\vect{m}, \vect{m}^{\prime})$}
	\underline{and
		the Lagrangian ${L}_{\scriptscriptstyle M,k}( \vect{m}, \vect{m}^{\prime} )$}}

The function
$\bar{F}_{\scriptscriptstyle M}(\vect{m}, \vect{m}^{\prime} )$, which is the main constituent of 
${L}_{\scriptscriptstyle M,k}( \vect{m}, \vect{m}^{\prime} )$, 
is clearly 
non-differentiable in classic sense,
and hence  we need to use the notion of sub-gradient of 
$\bar{F}_{\scriptscriptstyle M}(\vect{m}, \vect{m}^{\prime} )$
in order to derive an iterative algorithm.  One of the main difference between sub-gradient and
gradient is that sub-gradient at a point may be non-unique and the set of sub-gradients is known
as sub-differntial \cite{boyd2004convex}. The sub-differential of 
$\bar{F}_{\scriptscriptstyle M}(\vect{m}, \vect{m}^{\prime} )$ at $\vect{m}$ is  denoted by 
$\partial_{\vect{m}} \bar{F}_{\scriptscriptstyle M}
(\vect{m}, \vect{m}^{\prime} )$ which is a subset of $\mathbb{R}^N$. We say $\vect{r} \in 
\partial_{\vect{m}} \bar{F}_{\scriptscriptstyle M}
(\vect{m}, \vect{m}^{\prime} ) $ if it satisfies the following
for any $\vect{m}_1$ in $\mathbb{R}^N$:
$$\bar{F}_{\scriptscriptstyle M}(\vect{m}_1, \vect{m}^{\prime} ) \ge 
\bar{F}_{\scriptscriptstyle M}(\vect{m}, \vect{m}^{\prime} ) + 
\vect{r}^T
(\vect{m}_1-\vect{m}).$$

Sub-gradient of a differentiable function is unique and is equal to the standard
derivative (gradient).  The sub-differentiation is linear under some mild conditions and hence, for
${L}_{\scriptscriptstyle M,k}(\vect{m}, \vect{m}^{\prime} )$
defined in the equation \eqref{eq:lmkdef2},  we have
\begin{align}
\medmath{\partial_{\vect{m}}   {L}_{\scriptscriptstyle M,k}(\vect{m}, \vect{m}^{\prime} )  =} &
\partial_{\vect{m}} \bar{F}_{\scriptscriptstyle M}(\vect{m}, \vect{m}^{\prime})
+
\nabla_{\vect{m}}   \frac{\beta}{2}\left\|\vect{m}-\bar{\vect{m}}\right\|^2_2\\
= & \partial_{\vect{m}}\bar{F}_{\scriptscriptstyle M}(\vect{m}, \vect{m}^{\prime})
+
\beta(\vect{m}- \bar{\vect{m}}).
\end{align}
The meaning of the above equation is that the set 
$\partial_{\vect{m}}   {L}_{\scriptscriptstyle M,k}(\vect{m}, \vect{m}^{\prime} )$
is obtained
by adding  $\beta(\vect{m}- \bar{\vect{m}})$ to every element of  the
set $\partial_{\vect{m}} \bar{F}_{\scriptscriptstyle M}(\vect{m}, \vect{m}^{\prime})$.
We will also need the concept of $\epsilon$-subdifferential of 
${L}_{\scriptscriptstyle M,k}(\vect{m}, \vect{m}^{\prime})$, denoted by  
$\partial_{\vect{m},\epsilon}   {L}_{\scriptscriptstyle M,k}(\vect{m}, \vect{m}^{\prime})$.
We say $\vect{r} \in \partial_{\vect{m},\epsilon}   {L}_{\scriptscriptstyle M,k}(\vect{m}, \vect{m}^{\prime})$ if it satisfies the following for any $\vect{m}_1 \in \mathbb{R}^N$:
$${L}_{\scriptscriptstyle M,k}(\vect{m}_1, \vect{m}^{\prime}) \ge 
{L}_{\scriptscriptstyle M,k}(\vect{m}, \vect{m}^{\prime}) + 
\vect{r}^T
(\vect{m}_1-\vect{m}) -  \epsilon,$$ where $\epsilon$ is given positive real number.

By using the first derivative of ${F}_{\scriptscriptstyle M}(\vect{m}, \vect{m}^{\prime})$  and its
approximation (\cref{eq:fmfd} and \cref{eq:fmfd2}),
we  need to find  a sub-gradient  and an $\epsilon$-sub-gradient for 
${L}_{\scriptscriptstyle M,k}(\vect{m},\vect{m}^{\prime})$.
Note that,    the gradient of ${F}_{\scriptscriptstyle M}(\vect{m}, \vect{m}^{\prime})+
\frac{\beta}{2}\left\|\vect{m}-\bar{\vect{m}}\right\|^2_2$ is given by 
$\pmb{\gamma}_{1}(\vect{m},\vect{m}^{\prime}) + \beta(\vect{m}-\bar{\vect{m}})$.
Interestingly,  this also becomes a sub-gradient of 
${L}_{\scriptscriptstyle M,k}( \vect{m},\vect{m}^{\prime} )=
\bar{F}_{\scriptscriptstyle M}(\vect{m}, \vect{m}^{\prime})+
\frac{\beta}{2}\left\|\vect{m}-\bar{\vect{m}}\right\|^2_2$, which is the main result of the following
proposition.
\begin{restatable}[Subgradient for iterations]{proposition}{iterations}
	\label{ch2:thm:prop1}
	The quantity  $$\pmb{\zeta}(\vect{m},\vect{m}^{\prime}) =
	\pmb{\gamma}_{1}(\vect{m},\vect{m}^{\prime}) + \beta(\vect{m}-\bar{\vect{m}})$$
	is a sub-gradient of ${L}_{\scriptscriptstyle M,k}( \vect{m},\vect{m}^{\prime} )$ for all $\vect{m} \in [l,u]^N$, and
	$\pmb{\zeta}_\Delta(\vect{m},\vect{m}^{\prime}) = \pmb{\gamma}_{1,\triangle}(\vect{m},\vect{m}^{\prime}) 
	+ \beta(\vect{m}-\bar{\vect{m}})$
	is an $\epsilon$-sub-gradient of ${L}_{\scriptscriptstyle M,k}( \vect{m},{\vect{m}}^{\prime} )$ for all $\vect{m} \in [l,u]^N$.
\end{restatable}

Note that the sub-gradient is a set,  and the expression given in the above  proposition, 
$\pmb{\zeta}(\vect{m},\vect{m}^{\prime})$ is one of sub-gradients of 
${L}_{\scriptscriptstyle M,k}
(\vect{m},\vect{m}^{\prime})$.   Note that for a given vector, $\vect{m}^*$,
to be a minimum,   at least one of the sub-gradients  has to be zero.    
In the following  proposition, we give an expression for a sub-gradient  that  has
to be zero at the point of local minimum $\vect{m}^*$.
\begin{restatable}[Subgradient for termination]{proposition}{termination}
	\label{ch2:prop2}
	The quantity 
	\begin{equation}   
	\label{ch2:eq:optsubdiff}
	\pmb{\eta}(\vect{m},\vect{m}^{\prime}) = 
	{\cal P}_{[l,u,\vect{m}]} (\pmb{\gamma}_1(\vect{m},\vect{m}^{\prime})+
	\beta(\vect{m}-\bar{\vect{m}})) 
	\end{equation}
	is a sub-gradient of ${L}_{\scriptscriptstyle M,k}( \vect{m}, \vect{m}^{\prime} )$
	that goes to zero at the minimum point, and   the quantity 
	\begin{equation}   
	\label{ch2:eq:optsubdiffappr}
	\pmb{\eta}_\Delta(\vect{m},\vect{m}^{\prime}) 
	= {\cal P}_{[l,u,\vect{m}]} (\pmb{\gamma}_{1,\Delta}(\vect{m},\vect{m}^{\prime})+
	\beta(\vect{m}-\bar{\vect{m}})) 
	\end{equation}
	goes to zero at the minimum point   as $\Delta \rightarrow \infty$,  where 
	${\cal P}_{[l,u,\vect{m}]} (\vect{x})$ is component-wise projection of $\vect{x}$ as per the
	following rule:  (i)   if $(\vect{m})_i = l$, then $(\vect{x})_i $ is projected onto the 
	non-positive real line; (ii)  if $(\vect{m})_i = u$, then $(\vect{x})_i $ is projected onto the 
	non-negative  real line; (ii)  if $l<(\vect{m})_i < u$, then $(\vect{x})_i$ is left
	unchanged.
\end{restatable}

\subsection{Iterative methods}
\subsubsection{Damped Newton iterations}
\label{sec45}
By using the $\epsilon$-sub-differential  given in the Proposition 1,  
$\pmb{\zeta}_\Delta(\vect{m},\vect{m'})$, \label{sec46}
we construct the following iteration for computing the minimum
of ${L}_{\scriptscriptstyle M,k}(\vect{m}, \vect{m}^{\prime})$, which finds the updated
estimate for the minimum,  $\vect{m}_{l+1}^{(k)}$, given the current estimate  
$\vect{m}_{l}^{(k)}$ with $l$
being the iteration index:
\begin{align}
\label{eq:inner}
&\vect{m}_{l+1}^{(k)}
=\\ \nonumber&\medmath{\mathlarger{\mathlarger{\cal P}}_{ {[ l,u]} }
	\left(
	\vect{m}_{l}^{(k)}-\alpha_l 
	\left[\pmb{\zeta}_{\Delta_l}( \vect{m}_{l}^{(k)},\vect{m}^{\prime})\right]\oslash 
	\left[ { \mathlarger{\Gamma_{l}}
		\left(  \pmb{\gamma}_{2,\triangle_l}(\vect{m}_{l}^{(k)},\vect{m}^{\prime}) + \beta\vect{1}\right)}
	\right]\right)}
\end{align}
In the above  proposed iteration, $ \Gamma_l(\cdot)$ is the projection of the argument
onto the  set$ [\frac{1}{\sqrt{\delta_l}},\sqrt{\delta_l}]$
where $\delta_l$  is an iteration dependent  positive number,
and $\Delta_l$ is the iterative dependent approximation width.
Note that $ \pmb{\gamma}_{2,\triangle_l}(\vect{m}_{l}^{(k)},\vect{m}^{\prime}) + \beta\vect{1}$
is the approximation for diagonal of
$\bar{\nabla}_{\vect{m}}^2({L}_{\scriptscriptstyle M,k}(\vect{m}, \vect{m}^{\prime}))$.
Now, we give the proposition guaranteeing the convergence of the above 
iteration, whose proof  is based on the convergence analysis of projected  
$\epsilon$-sub-gradient method of Bonnettini et al. \cite{bonettini2016scaling}.

\begin{restatable}[Convergence of Damped Newton iterations]{proposition}{DN}
	\label{ch2:prop4}
	If $\alpha_l=\frac{C}{l+1}$, $\Delta_l \rightarrow \infty$, and $\delta_l=1+\frac{C_2}{(l+1)^2}$,
	then iteration \eqref{eq:inner} converges to the minimum of the problem 
	given in the equation  \eqref{eq:lmkdef2},   where, 
	$C$ and $C_2$ are any two positive real  numbers.
\end{restatable}
The implication of the above proposition is that, this iterative method can be used
for $\mathlarger{\mathlarger{\cal I}}_{[\vect{m}]}[]$ introduced in the Section 
\ref{sec:admm},  and any termination tolerance can be attained because of the above
convergence statement.  This means that the required condition on the tolerance
that we will derive in the next subsection, can be met.

\subsubsection{Majorization-Minimization iteration}

Here we propose a majorization-minimization (MM) method for the simplified
cost function given below:\\
%  \begin{equation}
%\label{eq:auglagm3}
${L}^{\prime}_{\scriptscriptstyle M,k}(\vect{m},\vect{m}^{\prime})   =  
{F}_{\scriptscriptstyle M}(\vect{m},\vect{m}^{\prime}) + 
\frac{\beta}{2} \left\|\vect{m}-\bar{\vect{m}}^{(k+1)}\right\|^2_2.$   \\
%\end{equation}
Clearly, the difference now is that the we have used the actual log-likelihood
without the out-of-bound penalty.  The reason will be explained at the end
of this section.   The above function has no inter-dependance 
among the components of  the vectors $\vect{m}$,  $\vect{m}^{\prime}$
and $\bar{\vect{m}}^{(k+1)}$.  Hence the above cost can be written
pixel-wise,  and minimization can be carried out pixel-wise.
To this end,  we use $m$, $m^{\prime}$, and $\bar{m}$ to  
replace an individual component of $\vect{m}$,   $\vect{m}^{\prime}$ and
$\bar{\vect{m}}^{(k+1)}$.
With this  replacement, we have the following expression for the objective function:
%  \begin{equation}
$\label{eq:auglagmscalar}
{L}^{\prime}_{\scriptscriptstyle M,k}({m},{m}^{\prime})   =  
{F}_{\scriptscriptstyle M}({m},{m}^{\prime}) + 
\frac{\beta}{2} \left\|{m}-\bar{{m}}\right\|^2_2,  $ 
%\end{equation}

The proposed MM approach proceeds as 
follows. Let $m_l$ be the current estimate of the minimum of 
${L}^{\prime}_{\scriptscriptstyle M,k}( {m}, {m}^{\prime} )$,  and let
${G}_{\scriptscriptstyle M}({m}, {m}^{\prime}, m_l) $  be the
$m_l$-dependent auxiliary function, known as the surrogate (majorizing) function,  
satisfying   
\begin{eqnarray*}
	{G}_{\scriptscriptstyle M}({m}, {m}^{\prime}, m_l)  
	& = &  {F}_{\scriptscriptstyle M}({m}, {m}^{\prime}),\;\; 
	\mbox{if} \;\; m=m_l \\
	{G}_{\scriptscriptstyle M}({m}, {m}^{\prime}, m_l)  
	& > & 
	{F}_{\scriptscriptstyle M}({m}, {m}^{\prime})\;\;   \mbox{otherwise.}
\end{eqnarray*}
Then, given initialization $m_0$,  the iteration towards
computing the minimum of  
${L}^{\prime}_{\scriptscriptstyle M,k}( {m}, {m}^{\prime} )$
proceeds as follows until convergence:
\begin{equation}
\label{eq:admmmmm}
m_{l+1} = \argmin_{m} {G}_{\scriptscriptstyle M}({m}, {m}^{\prime}, m_l) +
\frac{\beta}{2} \left\|{m}-\bar{{m}}\right\|^2_2
\end{equation}
The function ${G}_{\scriptscriptstyle M}({m}, {m}^{\prime}, m_l) $ is called the mojorizer
of ${F}_{\scriptscriptstyle M}({m}, {m}^{\prime})$.

To get the majorizer for ${F}_{\scriptscriptstyle M}({m}, {m}^{\prime})$,  
we use the ideas from Expectation-Maximization (EM) methods \cite{wu1983convergence},
which are well-known for computing maximum likelihood estimates.
EM methods find series of lower-bounding functions by expectation operation,   and maximize
these functions to get the required MLE.  When,  we consider the negative of log-likelihood,
this is equivalent to finding a series of upper-bounding functions by expectation operation
and minimize them to  get the required MLE.  In our case, we do not directly minimize these functions, 
but we use them in the equation \eqref{eq:admmmmm}.
So far, this approach has been used to compute to the noise
parameters (e.g. $\alpha$ and $\sigma$) for the mixed Poisson-Gaussian noise model 
\cite{jezierska2014approach},  but we use here to solve the data-fitting subproblem of
the proposed ADMM method.

Let $p(m)$ and $w$ be Poisson random variable with mean $m$ corresponding to photon count 
and Gaussian random variable.  Now, Eq. \eqref{eq:pgform} means 
\begin{align} \label{hid_data}
m^{\prime}=\alpha {p}(m) +{w}.
\end{align}
Here, ${p}$  and ${w}$ are the hidden (latent) data. $P$ and $M^{\prime}
$ are the random variables used for denoting ${p}$ and $m^{\prime}$. 
The EM approach requires the expectation to be computed with respect to  the conditional density
$f_{P|M^{\prime},m}({p}|m^{\prime},m)$. So, we first write the expression for the density 
$f_{P|M^{\prime},m}({p}|m^{\prime},m)$:
$
f_{P|M^{\prime},m}(p|m^{\prime},m) =\frac{f_{P,M^{\prime}|m}
	(p,m^{\prime}|m)}{\sum_{{p}=0}^{\infty}f_{P,M^{\prime}|m}({p},m^{\prime}|m)}
$
Here, the above equation follows from the definition of conditional density and the
denominator is the marginal density 
$f_{M^{\prime}|m}(m^{\prime}|m)=\sum_{p=0}^{\infty}f_{P,M^{\prime}|m}(p,m^{\prime}|m)$. 
For the computation of the numerator, 
we need 
\begin{equation}
\label{eq:fpmpbm}
f_{P,M^{\prime}|m}(p,m^{\prime}|m)= 
f_{M^{\prime}|P,m}(m^{\prime}|p,m)f_{P|m}(p|m),
\end{equation}
where
$f_{M^{\prime}|P,m}(m^{\prime}|p,m)$ is Gaussian density because $w$ is a Gaussian random variable;
hence , $f_{M^{\prime}|P,m}(m^{\prime}|p,m)
=\frac{1}{\sqrt{2 \pi}\sigma}\exp(\frac{-(m^{\prime}-\alpha p)^2}{2 \sigma ^2})$ and 
$f_{P|m}(p|m)$ is Poisson with mean $m$ so $f_{P|m}(p|m)=e^{-m} \frac{m^p}{p!}$.
Substituting the expressions gives
\begin{equation*}       
f_{P|M^{\prime},m}(p|m^{\prime},m)
=\frac{{m^{p}\exp(-(\frac{{m^{\prime}}-\alpha {p}}{\sqrt{2}\sigma})^2-(m))}/{{p}!}}
{\sum_{{p}=0}^{\infty}m^{{p}}\exp(-(\frac{{m^{\prime}}-\alpha {p}}{\sqrt{2}\sigma})^2-(m))/{{p}!}}
\end{equation*}

Now, it can be shown that, for any given $m_l$,  expectation of 
$-\ln\;f_{P,M^{\prime}|m}(p,m^{\prime}|m)$   with respect  $f_{P|M^{\prime},m}(p|m^{\prime},m_l)$
majorizes  ${F}_{\scriptscriptstyle M}({m}, {m}^{\prime})$ which is the negative log-likelihood
of $m$ being the source of $m^{\prime}$ via the equation \eqref{hid_data}. Hence the 
$m_l$-dependent majorizing function  for ${F}_{\scriptscriptstyle M}({m}, {m}^{\prime})$
denoted by ${G}_{\scriptscriptstyle M}({m}, {m}^{\prime}, m_l)$ is given by
\begin{equation}
\label{eq:admmmem}
{G}_{\scriptscriptstyle M}({m}, {m}^{\prime}, m_l) = 
{\cal E}_{P|M^{\prime},m_l}(-\ln f_{P,M^{\prime}|m}(p,m^{\prime}|m))
\end{equation}
Substituting the above equation in the iterative minimization depicted in the equation
\eqref{eq:admmmmm}  give the required MM iteration, which is given the form of proposition below.

\begin{restatable}[MM iterations]{proposition}{MM}
	\label{ch2:prop5}
	The iteration specified by the equation  \eqref{eq:admmmmm}  with  
	${G}_{\scriptscriptstyle M}({m}, {m}^{\prime}, m_l)$ given by  the equation \eqref{eq:admmmem}
	can be expressed as
	\begin{equation}
	\label{eq:admmmemit}
	{(m)_{(l+1)}}=\frac{{\beta\bar{m}}-1+\sqrt{(\beta{\bar{m}}-1)^2+4{q}_l\beta}}{2\beta},
	\end{equation} 
	where ${q}_l=\mathcal E _{\vect{P}|M^{\prime},m_l}({{p}}|m^{\prime},m_l).$
\end{restatable}

When compared with the damped-Newton iterative scheme given in \cref{sec46},
we have an advantage that the sequence of iterates $\{m_l\}$ is guaranteed to be
positive.  This is the reason why we eliminated the out-of-bound penalty,  and 
used ${F}_{\scriptscriptstyle M}({m}, {m}^{\prime})$ as opposed to the  
damped-Newton method, which was built using $\bar{F}_{\scriptscriptstyle M}({m}, {m}^{\prime})$.
We compute  ${q}_l=\mathcal{E}_{\vect{P}|M^{\prime},m_l}({{p}}|m^{\prime},m_l)$  as given 
in \cite{jezierska2014approach}, which uses the approximation  similar the one
used in the equation \eqref{eq:sexp2}.  Because of this approximation,  the theoretical convergence properties of the algorithm are not known;   however, we observed that, in our experiments,
the above iteration always converged  with this approximation.

\subsection{Modified termination conditions for iterations}

Recall that, the modified  ADMM  framework of Eckstein and Yao
\cite{eckstein2017approximate}   handles the case where the inner
sub-problems cannot be solved exactly.  It states two alternative
conditions to be satisfied by inner iterations for ensuring overall 
convergence of ADMM iteration.  These conditions are state in 
Section \ref{sec:admmpgch}. These conditions, however, are not
suitable for our problem,  because, they are expressed in terms of
exact sub-gradients.   However, we cannot compute exact sub-gradients
of $L_{\scriptscriptstyle M, k}(\vect{m},  \vect{m}^{\prime})$ because
of the infinite summations, and we have only approximate 
sub-gradients. The conditions need to be modified such that overall
ADMM method converges if these approximate quantities are used instead
of the exact ones. Then the following proposition gives the modified
conditions.
\begin{restatable}[Sufficient conditions for convergence]{proposition}{sufficient}
	\label{ch2:prop3}
	The condition \\ $\|\pmb{\eta}^{(k)}_{l,\Delta_l}\| + \sqrt{N}\bar{B}(1-erf(\Delta_l/\sqrt{2}))
	< \theta_k$ is sufficient for $\|\pmb{\eta}^{(k)}_l\| < \theta_k$ where $\bar{B}$ is 
	a constant that can be computed from parameters of the noise model and $\Delta_l$ is the  iteration dependent approximation width.
	The condition $\frac{a_{k,l}}{(c_k +  \|\vect{Hg}^{(k+1)}-\vect{m}^{(k)}_l\| ^2)} <  \rho <1$
	where $a_{k,l} = 2|\langle \vect{w}_{\Delta_l}^{(k)}-\vect{m}^{(k)}_{l},
	{\vect{\eta}_{l,\Delta_l}^{(k)}} \rangle|+(\|\vect{\eta}_{l,\Delta_l}^{(k)}
	\|+\|e_l\|)^2+Z_l(w_{\Delta_l}^{(k)},\vect{\eta}_{l,\Delta_l}^{(k)}) $
	is sufficient for $\frac{2|\langle\vect{w}^{(k)}-\vect{m}^{(k)}_l,
		\vect{\pmb{\eta}^{(k)}_l}\rangle|+\|\pmb{\eta}^{(k)}_l\| ^2}
	{c_k+ \|\vect{Hg}^{(k+1)}-\vect{m}^{(k)}_l\| ^2}<\rho<1$. \\Here,  $Z_l(w_{\Delta_l}^{(k)},\vect{\eta}_{l,\Delta_l}^{(k)})=2\|\vect{w}_{\Delta}^{(k)}-\vect{m}^{(k)}_{l}\| \  \|{e_l}\|+(                       2\beta\sum_{i=0}^{k-1} \|e_i\| ) \  (\|\eta_{\Delta_l}{(\vect{m}^{(k)}_{l},\vect{m}')} \|+\|e_l\| )$ \\and $\|e_j\|  \leq \sqrt{N}\bar{B} (1-erf(\frac{\Delta_j}{\sqrt{2}})) .$
\end{restatable}
The implications of the above result is that, in addition to the fact that 
the number of iteration in $\mathlarger{\mathlarger{\cal I}}_{[\vect{m}]}
[L_{\scriptscriptstyle M,k}(\cdot), \cdot, \cdot]$ should increase as 
$k$ increases,  the approximation  width  $\Delta_k$ should also increase
as $k$ increases.   

%%%%% n^* definiton
%%%%% \bar{B} in prop 5

\section{Experimental results}
\label{sec:majhead}

\begin{figure}
	\begin{minipage}[t]{0.15\linewidth}
		\centering
		\includegraphics[width=\linewidth]{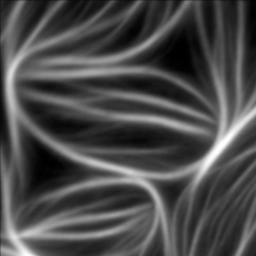}
		%	(a) Image1
	\end{minipage}%
	\vspace{.5mm}
	%	\hfill\vrule\hfill
	\begin{minipage}[t]{0.15\linewidth}
		\centering
		\includegraphics[width=\linewidth]{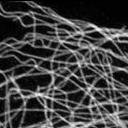}
	\end{minipage}
	\begin{minipage}[t]{0.15\linewidth}
		\centering
		\includegraphics[width=\linewidth]{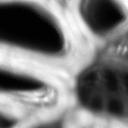}
		% (c) Image 3
	\end{minipage}
	\begin{minipage}[t]{0.15\linewidth}
		\centering
		\includegraphics[width=\linewidth]{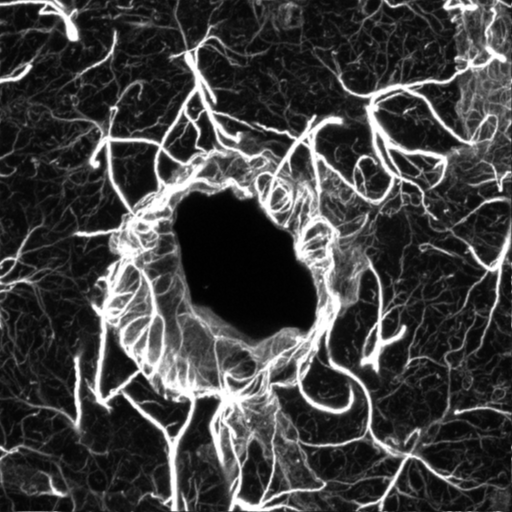}
		%	(a) Image1
	\end{minipage}%
	\vspace{.5mm}
	%	\hfill\vrule\hfill
	\begin{minipage}[t]{0.15\linewidth}
		\centering
		\includegraphics[width=\linewidth]{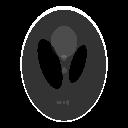}
	\end{minipage}
	\begin{minipage}[t]{0.15\linewidth}
		\centering
		\includegraphics[width=\linewidth]{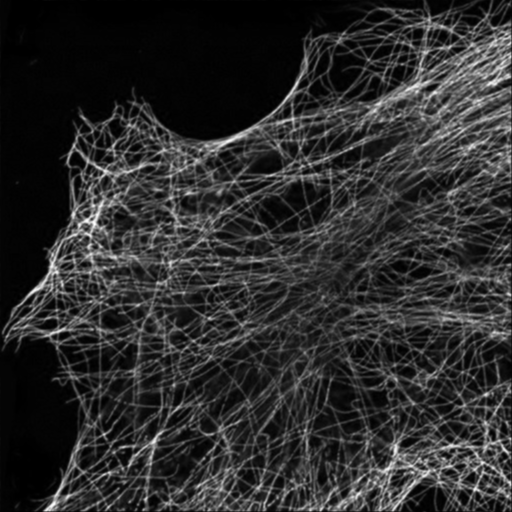}
		% (c) Image 3
	\end{minipage}
	\caption{\label{test_image}Images used for experiments}
\end{figure}

The proposed method is compared with primal-dual splitting method of
Chouzenoux et al. \cite{chouzenoux2015convex}
using de-blurring experiments. We consider six  images that are typical to 
fluorescence microscopy (e.g filament-like structures) as given in Figure \ref{test_image}.  To generate
the measured images,  we consider the following parametric form of the noise model:
\begin{equation} 
\label{eqscale}
(\vect{\vect{m}^{\prime}})_{n}  = \alpha \mathcal P((\alpha'\vect{Hg})_{n})+ \mathcal N(0,\sigma^2),
\end{equation}
where $\mathcal P(\cdot)$  where  $\mathcal N(\cdot,\cdot)$ represent Poisson and Gaussian noise
processes.  Here,   $\alpha^{\prime}$ serves as control  for the product of the exposure time and
intensity of the signal hitting the acquisition device, which is directly proportional to the excitation light intensity.
Although this scale factor can be  absorbed into  the image $\vect{g}$, the above representation helps to 
study the effect of varying the exposure time and excitation intensity.  The factor $\alpha$ determines the
efficiency of converting the detected photons into electron as well as a possible amplification that can be
applied on the detected electrons.  
For deblurring each noisy image,  we use the same regularization weight
($\lambda$) and boundary conditions (periodic) in both methods. Regularization parameter is chosen to
be  the lowest  value required to eliminate noise and noise-related artifacts. 
This lowest value is heuristically determined by using a series of trials involving small steps
with  a starting  value that is sufficiently low.
We observed that it was not required to tune  the step size, $\alpha_l$
\cref{eq:inner},  for each  
outer iteration index (ADMM) as well as inner iteration index in practice.
Also,  it was not required to tune it for each input image, and the  
value of $1.0$ worked for all  images.
Also, $\beta$ was fixed to 1 for all simulations.
The simulations are carried out on Intel Core i7-2600 CPU with 3.40GHz and 16GB 
RAM running on Ubuntu 16.04. In order to compare the performance three sets of experiments were conducted.

\begin{figure}
	%\centering
	\includegraphics[width=\linewidth]{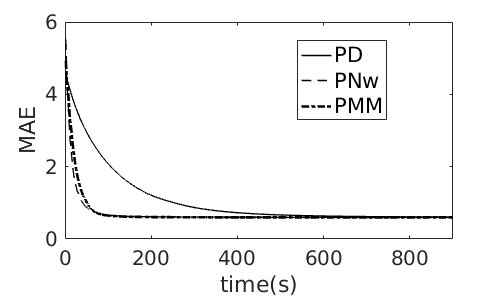}
	\caption{  Evolution of of the result w.r.t. time for Im 1 with  $\sigma=3$}
	\label{fig:evol}
	
\end{figure}

\begin{figure}[!htb]
	\caption{Partially restored Images for Im 1 $\sigma=3$ (TV)\label{snap}}			
	\centering
	\includegraphics[width=8.7cm]{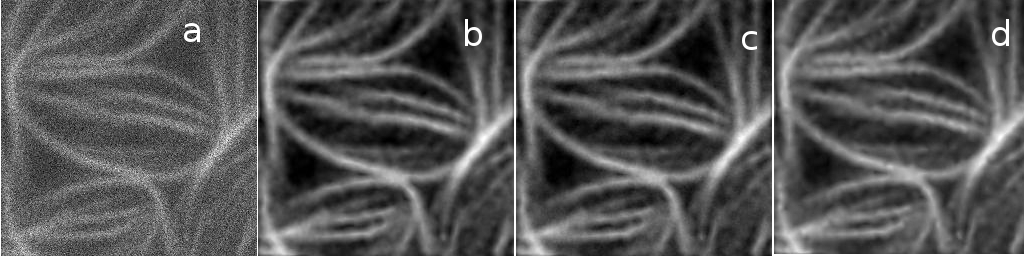}\\
	(a)Noisy     (b)PNw     (c) PMM     (d)PD
\end{figure}

\begin{figure}[ht]
	%\centering
	\includegraphics[width=\linewidth ]{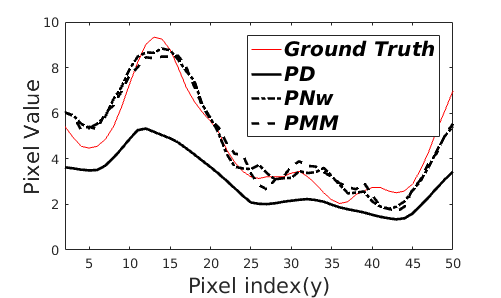}
	\caption{Scan lines for Im 1 $\sigma=3$ for \cref{snap}}
	\label{fig:scan}
\end{figure}

\begin{figure}
	\centering
	\includegraphics[width=.5\linewidth]{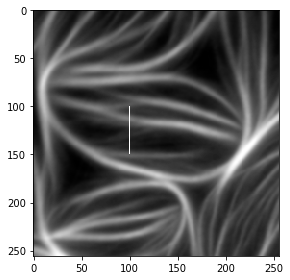}
	\caption{  Cross-section (vertical line) for scan lines in \cref{fig:scan}  }
	\label{fig:cross_sec}
	
\end{figure}

In the first set of experiments, we set $\vect{g}$ to have its maximum value in the
range $3-30$,   and set 
$\alpha^{\prime}$ to be $1$. The  images were blurred by realistic 2D TIRF PSF 
(Point Spread Function)  corresponding to 1.4 NA objective lens with  713 nm as the emission
wavelength    and with 133 nm as the sampling step size.
We used two different values for $\sigma$ and fixed $\alpha$ at $1$.  
This makes  a total of 12  test datasets.   Regularization was set to TV-2.
The results are displayed  in  \cref{big-tabel} ,  where we are comparing the mean absolute error (MAE)
attained by different methods after 100 and 200 seconds of  computation.  Note that,  since all  methods 
minimize the same cost function,  the final  MAE is the same for all methods.   It is clear from the comparisons
that the both variant of proposed method are significantly faster than the primal-dual method.  
It is also clear that the  proposed methods achieves the
final MAE  faster and the speed is less sensitive to variations in $\sigma$. Although, the primal-dual splitting method is slightly faster in case of Im1 and Im3 for larger values of $\sigma$,    the primal-dual splitting method is significantly slower for the lower values of $\sigma$.  We study in detail one test case (Im1, $\sigma=3$) from this set. Figure \ref{fig:evol} compares progression of MAE for both methods with respect to 
time towards the final solution 
for this deblurring trial. The figure clearly confirms 
that our methods converges faster. Figure \ref{snap} compares the snapshots of the 
methods at 100s for the same test case,  which also confirms that our method attains an  MAE close to  the
final MAE faster. Figure \ref{fig:scan} compares the scans line obtained the images of  \cref{snap}
for a closer view. These scans are obtained from a cross section shown 
as the white vertical line in \cref{fig:cross_sec}.

In the second set of experiments (\cref{scale}) the scale prior to the poisson process,
$\alpha^{\prime}$,  is varied.  Recall that we set $\alpha^{\prime}=1$ in the previous
experiment, and here we consider two additional values  this parameters, i.e,  we we  set
$\alpha^{\prime} = 0.75, 2$.  To add variety in terms of the algebraic structure of the cost,  we also consider two additional PSFs:   one with  emission wavelength of 
650 nm  and step size of 64 nm (for Im3), and another with  emission wavelength of
680 nm and step size 64 nm (for Im6).   Numerical Aperture  for all the PSFs was set to 1.41. In \cref{scale}, MAE at various time instances are compared for the two algorithms to investigate the sensitivity of the algorithms to the scale of the input. As in the previous experiments, the proposed methods are much less sensitive to the variation in scale. Also, the proposed methods  is uniformly faster than the PD  method except for Im3 
with  $\alpha'=.75$, in which is case PD method is slightly faster. However, in this case,
difference in speed is insignificant. On the other hand, the proposed   methods are
much faster  than PD method for $\alpha'=2$.

\begin{table*}[h]
	\centering
	\caption{Comparison of MAE at various execution times; PD: primal-dual splitting method  \cite{chouzenoux2015convex};
		PNw: proposed method with damped-Newton inner iteration; PMM:  proposed method with MM based inner iteration.}
	\label{big-tabel}
	\begin{tabular}{||l|l||l|l|l|l|l|l|l||}
		\hline
		Im              & $\sigma$ & \begin{tabular}[c]{@{}l@{}}Final\\ (all)\end{tabular} & PNw (200s) & PMM (200s) & PD (200s) & PNw (100s) & PMM (100s) & PD (100s) \\ \hline \hline
		\multirow{2}{*}{1} & 3                     & 0.598                                                 & 0.609     & 0.610     & 1.114    & 0.624     & 0.668     & 1.961    \\  
		& 4                     & 0.681                                                 & 0.702     & 0.735     & 0.681    & 0.819     & 1.167     & 0.687    \\ \hline
		\multirow{2}{*}{2} & 2.5                   & 1.759                                                 & 1.760     & 1.759     & 3.307    & 1.760     & 1.761     & 4.148    \\  
		& 4                     & 1.990                                                 & 1.990     & 1.991     & 1.990    & 1.991     & 1.990     & 1.998    \\ \hline
		\multirow{2}{*}{3} & 3                     & 0.959                                                 & 0.959     & 0.959     & 8.820    & 0.979     & 0.965     & 10.762   \\ 
		& 4                     & 1.029                                                 & 1.029     & 1.031     & 1.029    & 1.031     & 1.050     & 1.029    \\ \hline
		\multirow{2}{*}{4} & 1                     & 0.308                                                 & 0.308     & 0.308     & 1.0432   & 0.311     & 0.311     & 1.0432   \\  
		& 2                     & 0.351                                                 & 0.354     & 0.362     & 0.352    & 0.360     & 0.377     & 0.359    \\ \hline
		\multirow{2}{*}{5} & 2                     & 0.551                                                 & 0.551     & 0.559     & 0.609    & 0.554     & 0.566     & 0.749    \\  
		& 2.5                   & 0.971                                                 & 0.971     & 0.981     & 1.876    & 0.977     & 0.987     & 2.172    \\ \hline
		\multirow{2}{*}{6} & 2.5                   & 1.642                                                 & 1.646     & 1.653     & 4.451    & 1.665     & 1.685     & 4.502    \\ 
		& 3                     & 1.694                                                 & 1.699     & 1.710     & 2.485    & 1.718     & 1.768     & 3.078    \\ \hline \hline
	\end{tabular}
\end{table*}

\begin{table*}[]
	\caption{Comparison of MAE at various execution times; PD: primal-dual splitting method  for different input scales ;
		PNw: proposed method with damped-Newton inner iteration; PMM:  proposed method with MM based inner iteration.}
	\label{scale}
	\centering
	\begin{tabular}{||l|l||l|l|l|l|l|l|l|}
		\hline
		& scale($\alpha'$) & Final-all   & PNw 
		(200s) & PMM 
		(200s) & PD 
		(200s) & PNw 
		(100s) & PMM 
		(100s) & PD 
		(100s) \\ \hline \hline
		\multirow{2}{*}{Im1} & 0.75              & 0.503 & 0.511   & 0.514   & 0.716  & 0.553   & 0.551   & 1.456  \\ 
		& 2                & 0.915 & 0.951   & 0.943   & 11.331 & 1.208   & 1.229   & 11.506 \\ \hline
		\multirow{2}{*}{Im3} & 0.75              & 1.202 & 1.226   & 1.2205  & 1.1932 & 1.268   & 1.270   & 1.189  \\ 
		& 2                & 3.560 & 3.590   & 3.599   & 8.929  & 3.654   & 3.668   & 9.504  \\ \hline
		\multirow{2}{*}{Im6} & 0.75              & 0.853 & 0.853   & 0.854   & 4.158  & 0.865    & 0.863    & 5.800  \\ 
		& 2                & 1.643 & 1.647   & 1.647   & 27.66  & 1.938   & 2.002   & 27.64  \\ \cline{2-9} \hline  \hline
	\end{tabular}
\end{table*}

\begin{table*}[]
	\caption{Comparison of MAE at various execution times; PD: primal-dual splitting method  for Hessian-Schatten norm ;
		PNw: proposed method with damped-Newton inner iteration; PMM:  proposed method with MM based inner iteration.}
	\label{hs_table}
	\centering
	\begin{tabular}{||p{.60cm}|p{.65cm}||p{.85cm}|p{.75cm}|p{.75cm}|p{.75cm}|p{.75cm}|p{.75cm}|p{.75cm}||}
		\hline 
		Im              & $\sigma$ & \begin{tabular}[c]{@{}l@{}}Final\\ (all)\end{tabular} & PNw
		(200s) & PMM  
		(200s) & PD
		(200s) & PNw
		(100s) & PMM
		(100s) & PD
		(100s) \\ \hline \hline
		\multirow{2}{*}{1} & 3                     & 0.595                                                 & 0.613     & 0.610     & 1.882    & 0.686      & 0.689      & 2.777    \\ %\cline{2-9} 
		& 4                     & 0.683                                                 & .703      & 0.704      & 0.686     & 0.801      & 0.772      & 0.739     \\ \hline
		\multirow{2}{*}{2} & 2.5                   & 1.651                                                 & 1.653     & 1.652     & 4.143    & 1.655     & 1.654     & 4.932  \\ %\cline{2-9} 
		& 4                     & 1.857                                                 & 1.857     & 1.858     & 1.857    & 1.866     & 1.867     & 1.857    \\ \hline
		\multirow{2}{*}{3} & 3                     & 0.958                                                 & 0.956     & 0.955     & 11.384   & 0.962     & 0.970     & 11.486   \\ %\cline{2-9} 
		& 4                     & 1.028                                                 & 1.032     & 1.032     & 1.027    & 1.047     & 1.128     & 1.349    \\ \hline
		\multirow{2}{*}{4} & 1                     & 0.306                                                 & 0.308     & 0.308     & 1.0432   & 0.313     & 0.312     & 1.0432   \\ %\cline{2-9} 
		& 2                     & 0.348                                                 & 0.353     & 0.356     & 0.350    & 0.368     & 0.370     & 0.354    \\ \hline
		\multirow{2}{*}{5} & 2                     & 0.543                                                 & 0.548     & 0.550     & 0.578    & 0.560     & 0.607     & 0.836    \\ %\cline{2-9} 
		& 2.5                   & 0.917                                                 & 0.922     & 0.920     & 1.073    & 0.932     & 0.928     & 1.105    \\ \hline
		\multirow{2}{*}{6} & 2.5                   & 1.649                                                 & 1.665     & 1.667     & 5.019    & 1.701     & 1.725     & 5.135    \\ %\cline{2-9} 
		& 3                     & 1.698                                                 & 1.722     & 1.732     & 2.568    & 1.762     & 1.828     & 3.322    \\ \hline \hline
	\end{tabular}
\end{table*}
% Please add the following required packages to your document preamble:
% \usepackage{multirow}

\begin{table}[]
	\centering
	\caption{Number of gradient computations from experiment set 1.}
	\label{my-label}
	\begin{tabular}{|l|l|l|l|l|l|}
		\hline
		Im & $\sigma$ & Target  MAE & PNw & PMM  & PD \\ \hline
		1     &  3     & 1             & 52     & 154 &3940    \\ \hline
		1 &  4  &  1           &     74    & 261    & 500    \\ \hline
		3 & 4       &  1.5            & 183       & 480    &   2235 \\ \hline
		5 & 2       &  1            & 4      & 15    &   2250 \\ \hline
		6 & 3       &  2            &34       & 112    &   5060 \\ \hline
	\end{tabular}
\end{table}



\begin{figure}[ht]
	\caption{Partially restored Images for Im2 $\sigma=2.5$ at 100s (Hessian Schatten) \label{snap2}}	
	\centering
	\includegraphics[width=8.7cm]{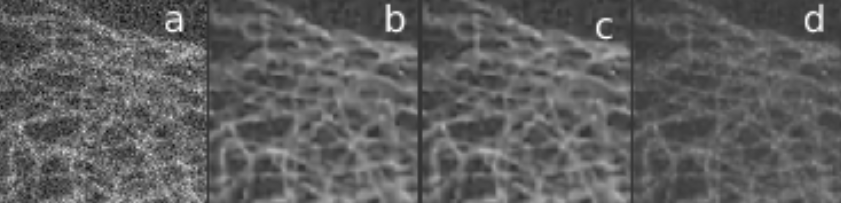}\\
	(a)Noisy     (b)PNw     (c) PMM     (d)PD
\end{figure}

In the third set of experiments,  the test case of first set of experiment were rerun 
with  TV-2 regularization replaced by Hessian-Schatten regularization with $q=1$
(see \cref{eq:fddef1}).  The results are given in the  \cref{hs_table}.  Here too, the
relative performances of different methods   in terms of convergence speed,
confer to the same pattern as that of the first set of experiments.   However,  the
actual  MAEs attained by the methods are lower here,  because 
Hessian-Schatten norm with $q=1$  has superior structure-preserving ability. 
As a special case, we observe MAEs obtained by the PD method from data set 
simulated from Im 4 with $\sigma = 1$  obtained at 100s and 200s are 
identical.  This Means that PD method is converging very slowly because of high value
of Lipschitz constant.   Next, the time-snap shots
of partial results of  various optimization methods  applied on noisy-blurred 
image obtained  from Im2 with $\sigma = 2.5$  at 100s are given in the  
\cref{snap2}. It is clear from the displayed images that  results of proposed methods 
are visually better than the PD method.  Further,  similar to  the  first experiment 
set,    the primal-dual splitting algorithm is more sensitive to change in $\sigma$.

In \cref{my-label}, we show the number of gradient evaluations for the 
specified Target MAE corresponding to  various test cases from the experiment
set 1.  Although, we have shown the comparisons only for specific test cases,  we observed 
similar patterns for the entire data-set. These results 
show that primal-dual splitting method requires much more gradient evaluations which 
are expensive in this problem.  The results from the sets of experiments confirm that the speed of 
primal-dual splitting method is sensitive to $\sigma$, and the maximum value of the pixels
in blurred noisy image; this is  because the upper bound on step size is the inverse of the Lipschitz 
constant of the gradient of data likelihood,  which is proportional to 
$(1-e^{\frac{-1}{\sigma^2}}) \exp(\frac{2\max_i(\vect{y}_i)-1}{\sigma^2}) $
\cite{chouzenoux2015convex}.  
This limitation is clearly not present in the proposed methods.  
Hence, the proposed algorithm has  a wider applicability.

\section {Conclusions}

We developed an ADMM based computational method for image restoration
under mixed Poisson-Gaussian (PG)  noise using  convex non-differentiable 
regularization functionals.  The main challenge was that there are no known
methods for computing the proximal solution of PG log-likelihood functional,
which is required   for adopting ADMM approach for this problem.  We 
developed  iterative methods for computing the proximal solution of the 
PG log-likelihood functional  along with the derivation of convergence proof.
We also derived termination conditions for these iterative methods to be met
for using them inside the ADMM iterative loop.  This led to the first ADMM
based method for  image restoration under PG noise model using convex
non-differentiable regularization functional.  As in other image restoration
problems, here too we demonstrated that the ADMM based method is faster
than primal-dual splitting method.   It should be emphasized that 
the approach used for the proofs of convergence are general, and hence
the proposed method can be extended to  any other complex likelihood models provided that the model has an uniform approximation for the gradient. 

\section*{Appendix A: Proximal solutions for ADMM step}

Since, some of the minimizations involved in the above iterative procedure can be solved 
exactly and yield a closed form solution, we first explain these exact minimizations involved
in the procedure.
In the ADMM scheme given in the equations \eqref{eq:admmg2}-\eqref{eq:admmb2}, 
the quadratic minimization
problem given  in the  Step 1 (equations \eqref{eq:admmg2} and \eqref{eq:qdef}) can
obviously be solved exactly, because  equating the gradient of $Q(\cdot)$ gives a linear system
of equations,  whose solution can be expressed as follows:
\begin{align} \label{eq12} 
\vect{g}^{(k+1)} & = (\vect{I}+\vect{D^TD}+\vect{H^TH})^{-1}\\ &  \medmath{\left[\vect{b}^{(k)}
	+\frac{1}{\beta}\hat{\vect{b}}^{(k)}
	+\vect{D^T}(\vect{d}^{(k)}+\frac{1}{\beta}\hat{\vect{d}}^{(k)}) \right. +  \left.   \vect{H^T(m}^{(k)}+\frac{1}{\beta}\hat{\vect{m}}^{(k)}  )\right]. }
\end{align}
As all matrices involved in the above computation are block circulant with circulant blocks 
(corresponding to 2-D circular convolution), the matrix inversion involved in the above step 
can be efficiently computed using FFTs. Similarly, efficient inversions can be done using DCT 
and DST for symmetric and anti-symmetric  boundary conditions respectively 
\cite{martucci1994symmetric}. Note that, even if the matrices involved are not block circulant,  the above inversion can be performed  iteratively and still the convergence holds as the framework of Eckstein et al. \cite{eckstein2017approximate} allows both the proximal operators to be inexact. 
Next, consider the subproblem corresponding to Step 3 (equation \eqref{eq:admmd2}).  Although
the sub-function is non-quadratic and non-differentiable,  solving the problem exactly is possible
thanks to the specific structure of the sub-function.
The solution is the well-known multidimensional shrinkage operation (\cite{xie2014admm}, eq. 3.24).
For the reader's convenience, we  express the solution for our notations.   
We first note that, as far as the minimization with respect to  $\vect{d}$ is concerned,
$L_{\scriptscriptstyle D}({\bf g}^{(k+1)},\vect{d},\hat{\vect{d}}^{(k)})$ in the equation \eqref{eq:auglagd} 
can be replaced by the following function, which differs from 
$L_{\scriptscriptstyle D}({\bf g}^{(k+1)},\vect{d},\hat{\vect{d}}^{(k)})$
only by a constant that is independent of $\vect{d}$:
\begin{equation}
\label{eq:auglagd2}
\begin{split}
{L}_{\scriptscriptstyle D,k}( \vect{d} )   =  & \medmath{\lambda F_{\scriptscriptstyle D}(\vect{d}) + 
	\frac{\beta}{2} \left\|\vect{d}-\bar{\vect{d}}^{(k+1)}\right\|^2_2 }   
; \medmath{\bar{\vect{d}}^{(k+1)} = \vect{D}{\bf g}^{(k+1)}-\frac{1}{\beta}\hat{\vect{d}}^{(k)}}
\end{split}
\end{equation}
Hence the minimization problem \eqref{eq:admmd2} can be expressed 
using the equation \eqref{eq:auglagd2} as
%\begin{equation}
%\label{eq:admmd3}
$$\vect{d}^{(k+1)} = \argmin_{\vect{d}}\lambda \sum_{i=1}^N
\left\|\mathcal{E}(\vect{P}_i^T\vect{d})\right\|_q + 
\frac{\beta}{2} \left\|\vect{d}-\bar{\vect{d}}^{(k+1)}\right\|^2_2.$$\\
% \end{equation}
Let ${\cal S}([v_1,v_2,v_3]^T) = \tbtm{v_1}{v_2}{v_3}$. Then 
the solution to the above problem can be written
as 
$$\mathbf{d}^{(k+1)} = \sum_i^{N^2} \mathbf{P}_i
{\cal H_T}(\mathbf{P}_i^T\bar{\mathbf{d}}^{(k+1)}, 
\lambda /\beta, q),$$
where 
$${\cal H_T}(\mathbf{x}, t, q) = 
\begin{cases}
max(
\|{\cal S}(\mathbf{x})\|_F-t,0)\frac{\mathbf{x}}{\|\cal{S}(\mathbf{x})\|_F}, \;
\mbox{for}\;\; q=2 \\
{\cal S}^{-1}(|||{\cal S}(\mathbf{x})|||_t), \;\; \mbox{for}\;\; q =1,
\end{cases}
$$
with $|||\cdot|||_t$  denoting the operator that applies soft-thresholding
on the Eigen values of  its matrix argument and returns the resulting matrix.
%Because of the special structure of $\{\vect{P}_i\}$'s, the above cost can be written as
%\begin{equation}
%\label{eq:admmd4}
%\vect{d}^{(k+1)} = \argmin_{\vect{d}} \lambda \sum_{i=1}^N\left\|\vect{P}_i^T\vect{d}\right\|_2 + 
%\frac{\beta}{2} \sum_{i=1}^N \left\|\vect{P}_i^T\vect{d}-
%\vect{P}_i^T\bar{\vect{d}}^{(k+1)}\right\|^2_2.
%\end{equation}
%Since, the minimization is separable in $i$ we minimize w.r.t sub-block $\vect{z_i}$ which is defined as ${\vect{z}}_i :=\vect{P}_i^T{\vect{d}}$.
%%Since any vector $\vect{x}$ can be written as   $\vect{x} =  \sum_{i=1}^N  \vect{P}_i\vect{P}_i^T\vect{x}$,  
%%We can write $ \vect{d}^{(k+1)} =  \sum_{i=1}^N  \vect{P}_i {\vect{z}}_i^*$,  where
%\begin{equation}
%\label{eq:admmdp}
%{\vect{z}}_i^* = \argmin_{\vect{z}_i\in\mathbb{R}^3}  \lambda \left\|\vect{z}_i\right\|_2 +  
%\frac{\beta}{2} \left\|\vect{z}_i-\bar{\vect{z}}_i\right\|^2_2,
%\end{equation}
%where $\bar{\vect{z}}_i =\vect{P}_i^T\bar{\vect{d}}^{(k+1)}$.  The solution
%can be expressed as 
%\begin{equation}
%\label{eq:admmdps}
%{\vect{z}}_i^* = max\left(\left\|\bar{\vect{z}}_i\right\|-\frac{\lambda}{\beta},0\right)
%\frac{\bar{\vect{z}}_i}{\left\|\bar{\vect{z}}_i\right\|}
%\end{equation}
%After the above minimizations, $\vect{d}^{(k+1)}$ can be reconstructed as\\ $\vect{d}^{(k+1)}
%=\sum_{i=0}^N \vect{P}_i\vect{z}_i$.\\

Next, we consider the subproblem of the Step 4.  
Here too, as far as the minimization with respect to  $\vect{b}$ is concerned,
$L_{\scriptscriptstyle B}({\bf g}^{(k+1)},\vect{b},\hat{\vect{b}}^{(k)})$ in the equation
\eqref{eq:auglagb}  can be replaced by the following function, which differs from 
$L_{\scriptscriptstyle B}({\bf g}^{(k+1)},\vect{b},\hat{\vect{b}}^{(k)})$
only by a constant that is independent of $\vect{b}$:
\begin{equation}
\label{eq:auglagb2}
\begin{split}
{L}_{\scriptscriptstyle B,k}( \vect{b} )   =  &  F_{\scriptscriptstyle B}(\vect{b}) + 
\frac{\beta}{2} \left\|\vect{b}-\bar{\vect{b}}^{(k+1)}\right\|^2_2 ; \bar{\vect{b}}^{(k+1)}
= {\bf g}^{(k+1)}-\frac{1}{\beta}\hat{\vect{b}}^{(k)}
\end{split}
\end{equation}
Hence step 4  can be expressed as
%\begin{equation}
%\label{eq:admmb3}
$ \vect{b}^{(k+1)} = \argmin_{\vect{b}} F_{\scriptscriptstyle B}(\vect{b})  + 
\frac{\beta}{2} \left\|\vect{b}-\bar{\vect{b}}^{(k+1)}\right\|^2_2.$
%  \end{equation}
The solution to the above problem  can be expressed as \cite{bertsekas1999nonlinear}
$$\vect{b}^{(k+1)} = {\cal P}_{[0,u^{\prime}]}(\bar{\vect{b}}^{(k+1)})$$
where ${\cal P}_{[0,u^{\prime}]}(\cdot)$ is the projection of its argument
into the set bounded by the interval $[0,u^{\prime}]$. This projection
is essentially clipping  with the interval $[0,u^{\prime}]$.
\section*{Appendix B: Proof of propositions}
{\bf Proof of Proposition \ref{ch2:thm:prop1}   }:
Since the data fitting cost is separable across pixel indices,  we replace
$\vect{m}$ and $\vect{\bar{m}}$ by ${m}$ and $\bar{m}$ to denote
a chosen pixel.
Since,  $ \pmb{\gamma}_{1}({m},{m}^{\prime})$ is the gradient of
$F_{\scriptscriptstyle{M}}({m},{m'})$, it satisfies the inequality that any sub-gradient
should satisfy since the function is convex.
Next,   since $\bar{F}_{\scriptscriptstyle{M}}({m},{m'})$  is
$\infty$  when $m$ goes out of bounds,  $ \pmb{\gamma}_{1}({m},{m}^{\prime})$
should also satisfy the sub-gradient inequality for $\bar{F}_{\scriptscriptstyle{M}}({m},{m'})$ for $m \in [l,u]$.
This means that $\pmb{\gamma}_{1}({m},{m}^{\prime})+\beta(m-\bar{m})$
is a sub-gradient for $L_{M,k}(m,m^{\prime})$ for $m \in [l,u]$.
Now if we show that  $\pmb{\gamma}_{1,\Delta}({m},{m}^{\prime})$ is
$\epsilon$-sub-gradient of $F_{M}(m,m^{\prime})$,  
$\pmb{\gamma}_{1,\Delta}({m},{m}^{\prime})+\beta(m-m^{\prime})$  will be
clearly be the $\epsilon$-sub-gradient of $L_{M,k}(m,m^{\prime})$ for $m \in [l,u]$
because of the same set of arguments used above.  
To this end, we first  note that 
${F}_{\scriptscriptstyle M}(m_1,m') \geq{F}_{\scriptscriptstyle M}(m,m')
+\gamma_1(m,m')(m_1-m)$.
This can be written as \\
${F}_{\scriptscriptstyle M}(m_1,m') \geq{F}_{\scriptscriptstyle M}(m,m')
+\gamma_{1,\Delta}(m,m')(m_1-m)+(\gamma_1(m,m')-
\gamma_{1,\Delta}(m,m'))(m_1-m)$.
Next we show below that   $|\gamma_1(m,m')-
\gamma_{1,\Delta}(m,m')|\leq \bar{B}(1-erf(\frac{\Delta}{\sqrt{2}}))$
where $\bar{B}$ is a constant and $erf$ is error function \cite{erfc}.  
Since $erf(\frac{\Delta}{\sqrt{2}})\rightarrow 1$
as $\Delta\rightarrow \infty$,  we have that
${F}_{\scriptscriptstyle M}(m_1,m') \geq{F}_{\scriptscriptstyle M}(m,y)
+\gamma_{1,\Delta}(m,m')(m_1-m)-\epsilon$,
where $\epsilon=\bar{B}(1-erf(\frac{\Delta}{\sqrt{2}}))(u-l)$ is real number that goes 
to zero as $\Delta\rightarrow \infty$.
This means that  $\pmb{\gamma}_{1,\Delta}({m},{m}^{\prime})$ is an
$\epsilon$-sub-gradient of $F_{M}(m,m^{\prime})$.

\noindent
\textbf{To show that 
	$|\gamma_{1,\Delta}(m,m')-\gamma_{1}(m,m')|\leq \bar{B}(1-erf(\frac{\Delta}{\sqrt{2}}) )$
	where $\bar{B}$ is a constant}:\\
From the equations \cref{eq:fmfd} and \cref{eq:fmfd2},   we can deduce that
$|\gamma_{1,\Delta}(m,m')-\gamma_{1}(m,m')|= 
\frac{|s(m,m'-1)s_{\Delta}(m,m')-s_{\Delta}(m,m'-1)s(m,m')|}{s(m,m')s_{\Delta}(m,m')}.$
Next, from expressions  \eqref{eq:sexp} and \eqref{eq:sexp2}, we find that both
$s(m,m')$ and  $s_{\Delta}(m,m')$ are lower bounded by 
$\exp(-\frac{(m')^2}{2\sigma^2})$. Letting $B = \exp(\frac{(m')^2}{\sigma^2})$ gives \\
$|\gamma_{1,\Delta}(m,m')-\gamma_{1}(m,m')|
\le
B|s(m,m'-1)s_{\Delta}(m,m')-s_{\Delta}(m,m'-1)s(m,m')|.$
Adding and subtracting  $s(m,m')s(m,m'-1)$ gives
\begin{align}&|\gamma_{1,\Delta}(m,m')-\gamma_{1}(m,m')|
\leq\\ \nonumber& B\Big[s(m,m'-1)\bar{s}_{\Delta}(m,m')
+s(m,m')\bar{s}_{\Delta}(m,m'-1)\Big],\end{align}
where $\bar{s}_{\Delta}(m,m')=|s(m,m^{\prime})-s_{\Delta}(m,m^{\prime})|.$
Chouzenoux et al. \cite{chouzenoux2015convex} 
have shown that
$\bar{s}_{\Delta}(m,m') \leq \sqrt{2\pi}\sigma \frac{m^{n^*}}{n^*!} 
\exp(\frac{-(m'-\alpha n^*)^2}{2\sigma^2})(1-erf(\frac{\Delta}{\sqrt{2}}))$ for some $n^* \in \mathbb{N}$. Further note that, 
$\frac{m^{n^*}}{n^*!} \exp(\frac{-(m'-\alpha n^*)^2}{2\sigma^2}) \leq s(m,m')$.   Moreover,
note that $s(m,m')\leq \sum_{n=0}^\infty \frac{m^n}{n!}=e^m \leq e^u$, where $u$ is upper
bound on $m$.   Putting all these things together,  gives the required result.

\noindent	
{\bf{Proof of Proposition \ref{ch2:prop2}}}:
Here too, since the data fitting cost is separable across pixel indices,  we replace
$\vect{m}$ and $\vect{m}^{\prime}$ by ${m}$ and ${m}^{\prime}$ to denote
a chosen pixel.   We  have to show the following sub-gradient inequality 
for all $m$ in $[l,u]$ and $m_1\in \mathbb{R}$:\\
$L_{M,k}(m_1,m') \ge L_{M,k}(m,m') + \eta(m,m')(m_1-m).$\\
Since, $L_{M,k}(m_1,m')$ is $\infty$ for $m_1 \notin [l,u]$, and 
${\eta}({m},{m}^{\prime})$ is identical to $\zeta(m,m')$ for $m \in (l,u)$, it only
remains to show that ${\eta}({m},{m}^{\prime})$ satisfies the above equation
for $m$ in $\{l,u\}$ and $m_1\in [l,u]$.  Now for $m=l$,  ${\eta}({m},{m}^{\prime})$
is a projection of  $\zeta(m,m')$ onto non-positive real line, and hence,
${\zeta}({m},{m}^{\prime})\ge {\eta}({m},{m}^{\prime})$.  Since $(m_1-l)$ positive,
we have ${\zeta}({m},{m}^{\prime})(m_1-m)\ge {\eta}({m},{m}^{\prime})(m_1-m)$
for $m=l$.  In  a similar way, we can show that the above inequality is satisfied
for $m=u$ also.  This means that the sub-gradient inequality is satisfied for
$m$ in $\{l,u\}$ and $m_1\in [l,u]$.

It remains to be proven that $\eta(m,m') \rightarrow 0$ at minimum of $L_{M,k}$.
Note that finding minimum of  $L_{M,k}$ is equivalent to finding the minimum
of $L_{M,k}^{\prime}(m,m^{\prime}) 
= F_{M}(m,m^{\prime})+(\beta/2)(m-\bar{m})^2$ subject to
$m \in \Omega =  [l,u]$.  The first order necessary condition for the general case
is that the inner product between any feasible direction of the constraint set 
$\Omega$ \cite{bertsekas1999nonlinear} and the gradient should be non-negative.
In our problem,   for $m \in (l,u)$, the feasible directions
are both positive and negative real axes, and   hence, the first order condition means
that $\zeta(m,m^{\prime})=0$. Next, for $m = l$,  the feasible direction is positive real
axis, and hence,  the first order condition means that projection of $\zeta(m,m^{\prime})$
onto non-positive real axis should be zero. 
Further,  for $m = u$,  the feasible direction is negative real
axis, and hence,  the first order condition means that projection of $\zeta(m,m^{\prime})$
onto non-negative real axis should be zero. The above three statements imply
that $\eta(m,m^{\prime})$ should be zero.

Since, any projection is non expansive operator, we have the following:
$ |\gamma_1(m,m')-\gamma_{1,\Delta}(m,m')| \rightarrow 0$ as $\Delta \rightarrow \infty$, 
implies that  $|\eta(m,m')-\eta_{\Delta}(m,m')|   \rightarrow 0$ as $\Delta \rightarrow \infty$.
Hence  $\eta_{\Delta}(m,m')  \rightarrow 0$ as $\Delta \rightarrow \infty$ at the minimum
point.  

\noindent	
{\bf{Proof of Proposition \ref{ch2:prop4}}}:
We use the following theorem to prove  the proposition.
Consider iteration  $ x^{ (l+1)} = P _{X,D_l^{-1}}( x ^{(l)} 
-\alpha^{(l+1)}  D_l u^{(l)})$ for solving the problem
$\argmin_{\vect{x} \in X}f(\vect{x})$. 

Here, the scaled projection is defined as \\$P _{X,F}(\vect{g}) := \argmin_{x \in X}(\vect{(x-g)}^T F (\vect{(x-g)}))$ where $F$ is a positive definite matrix with bounded eigen values. $\vect{u}^
{(l)}$ is an $\epsilon_l$ sub-gradient of the $f$ at $\vect{x}^{(l)}$ i.e. $\vect{u}^
{(l)} \in \partial_{\epsilon_l}
f(\vect{x}
^{(l)}
)$ and $D_l$ is a gradient scaling matrix(symmetric positive definite with bounded eigen values).
\begin{theorem}(Bonettini et al., \cite{bonettini2016scaling})
	Let ${x^{(l)}} \in \Omega$ be the sequence generated by the above iteration ,
	for a given sequence $\{\epsilon_l\} \subset R$, $\epsilon_l \geq 0$. Assume that the set of the solutions of the above minimization problem $X^*$
	is
	non-empty and that there exists a positive constant $\rho$ such that $ \|\vect{u}
	^{(l)} \| \leq \rho$ and a sequence of
	positive numbers ${L_l}$ such that $ \|D_l \| \leq L_l,  \|{D_l}
	^{-1} \|
	\leq
	L_l$, with $1 \leq L_l  \leq L$ for some positive
	constant $L$, for all $l \geq 0$. If the following conditions holds, then the sequence generated by the iterations converge to a point in $X^*$
	%	\begin{enumerate}
	(1) $\epsilon_l \rightarrow 0$
	(2) $\sum\limits_{l=0}^{\infty} \alpha^{(l)}=\infty$
	(3) $\sum\limits_{l=0}^{\infty} (\alpha^{(l)})^2<\infty$
	(4) $\sum\limits_{l=0}^{\infty} \alpha^{(l)}  \epsilon_l<\infty$
	(5) $L_l^2=1+\gamma_l$,$\sum\limits_{l=0}^{\infty}\gamma_l < \infty$
	%	\end{enumerate}
\end{theorem}
In our case, the iterations proposed are,
\begin{align} \label{inner}
&\vect{m}_{l+1}^{(k)}
=\\&\medmath{ {\mathlarger{\cal P}}_{ {[ l,u]} }
	\left(
	{m}_{l}^{(k)}-\alpha_l 
	\left[{\zeta}_{1,\Delta_l}( {m}_{l}^{(k)},{m}^{\prime})\right]/
	\left[ { \mathlarger{\Gamma_{l}}
		\left(  {\gamma}_{2,\triangle_l}({m}_{l}^{(k)},{m}^{\prime}) + \beta\right)}
	\right]\right)}
\end{align}
Since, our set $X$ is $[m,M]$,  simply one dimension interval, scaled projection is same as simple projection.\\
Also our scaling factor is obtained after the projection on set$[\frac{1}{\sqrt{\delta_l}},\sqrt{\delta_l}]$ and $\delta_l$ which plays the role of $L_l^2$ is chosen to be  $\delta_l=1+\frac{C_2}{(l+1)^2}$, because of this scaling factor is bounded between $1$ and $\sqrt{1+C_2}$.\\
It has been proved previously in this paper that $\zeta_{\Delta_l}(m,m')$ is an $\epsilon$ Sub-gradient of $L_{M,k}(m,m',\bar{m})$ i.e. $\zeta_{\Delta_l}(m,m') \in$ $\partial _{\epsilon}{F_M(m,m')}$ where $\epsilon=\bar{B}(1-erf(\frac{\Delta}{\sqrt{2}}))|u-l|$.  Clearly, the $\epsilon$-sub-gradient is upper-bounded. Since $\Delta$ is the term that determines the number of terms in summation, for convergence we increase $\Delta_l$ in as iteration number increases by the following rule: $\Delta_l= [ C_{\Delta}l+1 ]$ (where $C_\Delta$ is any positive constant, $[x]$ denotes the greatest integer less than x).  Now since, $\Delta_l \rightarrow \infty \implies \bar{B}(1-erf(\frac{\Delta}{\sqrt{2}}))|u-l|\rightarrow 0$ as $erf(\frac{\Delta}{\sqrt{2}}) \rightarrow 1$. This justifies satisfiability of condition (1) of the theorem.\\
Since, $\alpha^{(l)}$ is chosen as  $\alpha^{(l)}=\frac{C}{l+1}$, conditions (2) and (3) are satisfies as it is a well known square summable sequence but not summable.\\
On substituting $\Delta_l= [ C_{\Delta}l+1 ]$, we get $\epsilon_l=\bar{B}(1-erf(\frac{[C_{\Delta}l+1]}{\sqrt{2}}))|l-u|$ 
\newline
Since, $1-erf(z)<\frac{exp(-z^2)}{\sqrt{\pi}z}
$ 
\cite{erfc}$\implies \epsilon_l<\sqrt{2} \bar{B}(\frac{\exp(-.5{[ C_{\Delta}l+1 ]}^2)}{\sqrt{\pi}[ C_{\Delta}l+1]})|u-l|  $ \\
Using $(\exp(-.5[C_{\Delta}l+1]^2))<1$ and substituting  for $\alpha^{(l)}$, we get,\\
$\sum\limits_{l=0}^{\infty} \alpha_l  \epsilon_l<\alpha_0\epsilon_0+\sum\limits_{l=1}^{\infty}\sqrt{2} \bar{B}\frac{C}{\sqrt{\pi} C_{\Delta}l(l+1)})|l-u|< \infty$.  This satisfies condition (4).

On comparing with our set of iterations $\delta_l=1+\frac{C_2}{(l+1)^2}$ plays the role of $L_l^2$, therefore 
$\gamma_l$ for our case becomes, $\frac{C_2}{(l+1)^2}$ which is a hummable sequence, and hence
this satisfies condition (5)

\noindent	
{\bf{Proof of Proposition \ref{ch2:prop5}}}:
From equation (\ref{eq:fpmpbm}),  we get 
$f_{\scriptscriptstyle P,M'|m}({p},{m'}|m)=\frac{m^p}{\sqrt{2\pi}\sigma{p}!}
\exp(-(\frac{\vect{m'}-\alpha {p}}{\sqrt{2}\sigma})^2-(m)).$\\
Applying log gives taking the required expectation, we get\\
$G_M(m,m',m_l)=\mathcal{E}_{\scriptscriptstyle {P}|{M'},m_{l}} \{\ln f_{\scriptscriptstyle P,M'|m}({p},{m'}|m)\}$ 
$= \ln(m) \mathcal{E}_{{P}|{M'},m_{l}}({p}) -m +cons,$\\
where 
$\mathcal{E}_{\scriptscriptstyle {P}|\vect{M'},m_{l}}({p})=\sum_{\vect{p}=0}^{\infty}{p} f_{P|M',m}({p}|{y},m_{l}) \label{exp}$.
Then the minimization given in the equation (\ref{eq:admmmmm})  is equivalent to solving the following equation: \\
$\frac{d G_M(m,m',m_l)}{dm}+\beta({m}-\bar{m})=0  \Longrightarrow
\frac{{q}}{m_{(l+1)}}-1-\beta({(m)_{(l+1)}}+{\bar{m}})=0$.
Solving this  equation gives required expression.     	   	

\noindent	
{\bf{ Proof of Proposition \ref{ch2:prop3}}}:
We have $|\eta_{\Delta} (m,m' )-{\eta{(m,m')}}| \leq |\gamma_{1}(m,m')-\gamma_{1,\Delta}(m,m')|\leq 
\bar{B} (1-erf(\frac{\Delta}{\sqrt{2}}))$, the above inequation follows from  section showing Proof of proposition 1.
So, instead of condition (1), \ref{condition1}, a verifiable condition that can be checked is  
\\
$|\eta_{\Delta_l}(m_{l},m')|+\bar{B}(1-erf(\frac{\Delta_l}{\sqrt{2}}))< \theta_k$

Following the above approach a similar condition to remedy the inexactness in the gradient can be derived for condition (2). 
It can be observed that $\vect{w}^{(k)}=\vect{w}^{(0)}-\beta\sum_{i=0}^{k-1}\eta(\vect{{m}^{(i)}},\vect{m'})$. $\vect{w}^{(0)}$ can be chosen to be zero vector without disturbing the criteria. Now, to derive a sufficient condition  to obtain an upperbound on \\$ 2|\langle\vect{w}^{(k)}-\vect{m}^{(k)}_{l},{\eta(\vect{m}^{(k)}_{l},\vect{m'})}\rangle|+ \|{\eta(\vect{m}^{(k)}_{l},\vect{m'})} \|^2$ \\
$ =2|\langle{-\beta\sum_{i=0}^{k-1}\eta(\vect{m}^{(i)},\vect{m'})-\vect{m}^{(k)}_{l},{\eta(\vect{m}^{(k)}_{l},\vect{m'})}\rangle}|+ \|{\eta(\vect{m}^{(k)}_{l},\vect{m}')} \|^2$.\\
Now, we do not have the exact value of the $\eta(\cdot)$ but we have a $\Delta$ approximation which is based on $\Delta$ gradient approximation.  Letting  $\eta(\vect{m}^{(k)}_{l},\vect{m'})=\eta_{\Delta_l}({\vect{m^{(l)}}},\vect{m'})+e_l$ where $e_l$ is the approximation error, gives\\
% $ {=2|\langle{c\sum_{i=0}^{k-1}\eta_{\Delta_i}({\vect{m^{(i)}}},\vect{m'})+\vect{m}^{(i)},{\gamma(\vect{m}^{(k)}_{l})}\rangle|+ \|{\eta({\vect{m^{(l)}}},\vect{m'})} \|^2}}$\\
$ \leq 2|\langle{\beta\sum_{i=0}^{k-1}\eta_{\Delta_i}({\vect{m}^{(i)},\vect{m'}})+e_i+\vect{m}^{(k)}_{l},{\eta(\vect{m}^{(k)}_{l},\vect{m'})}}\rangle|+ \|{\eta(\vect{m}^{(k)}_{l},\vect{m'})} \|^2$\\
$ \leq 2|\langle{\beta\sum_{i=0}^{k-1}\eta_{\Delta_i}({\vect{m}^{(i)},\vect{m'}})+\vect{m}^{(k)}_{l},\eta({\vect{m}^{(i)},\vect{m'} }})\rangle|+2\beta\sum_{i=0}^{k-1}| \langle e_i,\eta({\vect{m}^{(i)},\vect{m'}}) \rangle|   + \|\eta_{\Delta_l}({\vect{m}^{(k)}_{l},\vect{m'}})+e_l \|^2$ (Using Triangle inequality).\\
Now, substitute $\vect{w}_{\Delta}^{(k)}=-\beta\sum_{i=0}^{k-1}\eta_{\Delta_i}{(\vect{m}^{(i)},\vect{m'})}$\ and use cauchy Schwartz Inequality to get\\
\begin{align*} &= 2|\langle \vect{w}_{\Delta}^{(k)}-\vect{m}^{(k)}_{l},{\eta_{\Delta_l}(\vect{m}^{(k)}_{l},\vect{m'})}+e_l \rangle|+\\&2\beta\sum_{i=0}^{k-1}| \langle e_i,\eta_{\Delta_l}{(\vect{m}^{(k)}_{l},\vect{m'})}+e_l \rangle| + \|{\eta_{\Delta_l}(\vect{m}^{(k)}_{l},\vect{m'})}+e_l \|^2\\
&\le 2|\langle \vect{w}_{\Delta}^{(k)}-\vect{m}^{(k)}_{l},{\eta_{\Delta_l}(\vect{m}^{(k)}_{l},\vect{m'})} \rangle|\\&+2 \|\vect{w}_{\Delta}^{(k)}-\vect{m}^{(k)}_{l} \| \   \|{e_l} \|+( \|\eta_{\Delta_l}{(\vect{m}^{(k)}_{l},\vect{m'})}  \|+ \|e_l \| \\&                        +2\beta\sum_{i=0}^{k-1}  \| e_i \|) \  ( \|\eta_{\Delta_l}{(\vect{m}^{(k)}_{l},\vect{m}')}  \|+ \|e_l \|) \end{align*}
$=2|\langle \vect{w}_{\Delta}^{(k)}-\vect{m}^{(k)}_{l},{\vect{\eta}_{l,\Delta_l}^{(k)}} \rangle|+( \|\vect{\eta}_{l,\Delta_l}^{(k)}  \|+ \|e_l \|)^2+Z_l(w_{\Delta_l}^{(k)},\vect{\eta}_{l,\Delta_l}^{(k)})$, where
$Z_l=2 \|\vect{w}_{\Delta}^{(k)}-\vect{m}^{(k)}_{l} \| \   \|{e_l} \|+(                       2\beta\sum_{i=0}^{k-1}  \| e_i \|) \  ( \|\eta_{\Delta_l}{(\vect{m}^{(k)}_{l},\vect{m}')}  \|+ \|e_l \|)$.\\
Upperbound on each $ \|e_j \|$ can be obtained from as \\$ \|e_j \| \leq \sqrt{N}\bar{B} (1-erf(\frac{\Delta_j}{\sqrt{2}})) $ where N is the number of pixels in the image.
The above expression can be used to verify the required condition.\\

\bibliography{template.bib}
\bibliographystyle{IEEEtran}  
\end{document}